\documentclass[conference,compsoc]{IEEEtran}

\usepackage{pgf}
\usepackage{tikz}
\usepackage{subcaption}
\usepackage{filecontents}

\usepackage{booktabs}
\usepackage{amsmath}
\usepackage{cite}
\usepackage{todonotes}
\usepackage{url}
\usepackage{multirow}
\usepackage{tcolorbox}
\usepackage[x11names]{xcolor}
\usepackage[hidelinks]{hyperref}
\usepackage{amssymb}
\usepackage{float}
\usepackage{paralist}

\newcounter{recommendationcounter}
\renewcommand{\therecommendationcounter}{R\arabic{recommendationcounter}}
\newtcolorbox{recommendation}{
  colback=lightgray!30,
  notitle,
  boxsep=0pt,
  before upper={
        \refstepcounter{recommendationcounter}
        \textbf{\therecommendationcounter:}\quad 
    },
}
\newcounter{challengecounter}
\renewcommand{\thechallengecounter}{C\arabic{challengecounter}}
\newtcolorbox{challenge}{
  colback=darkgray!30,
  notitle,
  boxsep=0pt,
  before upper={
        \refstepcounter{challengecounter}
        \textbf{\thechallengecounter:}\quad
    },
}

\usepackage{xcolor}
\usepackage{fontawesome5}
\usepackage{enumitem}

\begin{document}

\title{\Large \bf SoK: The Last Line of Defense: On Backdoor Defense Evaluation}

\author{
\IEEEauthorblockN{Gorka Abad}
\IEEEauthorblockA{University of Bergen}
\and
\IEEEauthorblockN{Marina Krček}
\IEEEauthorblockA{Radboud University}
\and
\IEEEauthorblockN{Stefanos Koffas}
\IEEEauthorblockA{Delft University of Technology}   
\and
\IEEEauthorblockN{Behrad Tajalli}
\IEEEauthorblockA{Radboud University}   
\and
\IEEEauthorblockN{Marco Arazzi}
\IEEEauthorblockA{University of Pavia}   
\and
\IEEEauthorblockN{Roberto Riaño}
\IEEEauthorblockA{Ikerlan Research Center\\ Radboud University}   
\and
\IEEEauthorblockN{Xiaoyun Xu}
\IEEEauthorblockA{Radboud University}   
\and
\IEEEauthorblockN{Zhuoran Liu}
\IEEEauthorblockA{Radboud University}   
\and
\IEEEauthorblockN{Antonino Nocera}
\IEEEauthorblockA{University of Pavia}   
\and
\IEEEauthorblockN{Stjepan Picek}
\IEEEauthorblockA{Radboud University}   
}

\maketitle

\begin{abstract}
Backdoor attacks pose a significant threat to deep learning models by implanting hidden vulnerabilities that can be activated by malicious inputs. While numerous defenses have been proposed to mitigate these attacks, the heterogeneous landscape of evaluation methodologies hinders fair comparison between defenses.
This work presents a systematic (meta-)analysis of backdoor defenses through a comprehensive literature review and empirical evaluation. We analyzed 183 backdoor defense papers published between 2018 and 2025 across major AI and security venues, examining the properties and evaluation methodologies of these defenses. 

Our analysis reveals significant inconsistencies in experimental setups, evaluation metrics, and threat model assumptions in the literature. Through extensive experiments involving three datasets (MNIST, CIFAR-100, ImageNet-1K), four model architectures (ResNet-18, VGG-19, ViT-B/16, DenseNet-121), 16 representative defenses, and five commonly used attacks, totaling over 3\,000 experiments, we demonstrate that defense effectiveness varies substantially across different evaluation setups. 
We identify critical gaps in current evaluation practices, including insufficient reporting of computational overhead and behavior under benign conditions, bias in hyperparameter selection, and incomplete experimentation. Based on our findings, we provide concrete challenges and well-motivated recommendations to standardize and improve future defense evaluations. Our work aims to equip researchers and industry practitioners with actionable insights for developing, assessing, and deploying defenses to different systems.
\end{abstract}

\section{Introduction}
\label{sec:introduction}

Deep learning (DL) has become a primary technology in critical application domains, from autonomous vehicles~\cite{grigorescu2020survey} to medical diagnosis and malware classification~\cite{bensaoud2024survey,severi2021explanation}. As DL systems assume increasingly safety-critical roles, ensuring their security against adversarial manipulation has become increasingly important. Among the various threats to DL systems, including evasion attacks~\cite{szegedy2013intriguing}, model theft~\cite{jagielski2020high}, and data poisoning~\cite{jagielski2018manipulating}, backdoor attacks have emerged as particularly insidious and difficult to detect.

Backdoor attacks implant hidden malicious behavior into models that is activated only when specific trigger patterns appear in the input~\cite{gu2019badnets}. Unlike evasion attacks that occur at inference time, backdoors are embedded during training and persist throughout the model's lifetime, making them exceptionally dangerous for deployed systems. The attack surface is broad: adversaries can poison training data~\cite{gu2019badnets,chen2017targeted}, manipulate the training process, or directly modify model parameters~\cite{xu2025towards,rakin2020tbt}. Triggers can be embedded in the input space (e.g., pixel patterns~\cite{gu2019badnets}), the feature space (e.g., semantic perturbations~\cite{nguyen2021wanet,doan2021lira}), or the model weights, creating a diverse and evolving threat landscape.

The research community has responded with an extensive body of work on backdoor defenses, proposing detection and mitigation techniques that operate at different stages of the machine learning pipeline: pre-training (data sanitization), during training (robust learning), and post-training (model inspection)~\cite{li2022backdoor, wang2019neural, liu2019abs, xu2024ban}. This proliferation of defenses raises a critical question for practitioners: \textit{Which defense should be deployed in production systems?} Answering this question requires a rigorous, fair comparison of defense effectiveness under realistic threat models. However, our analysis reveals that the current state of defense evaluation makes such comparisons nearly impossible.

\textbf{The evaluation crisis.} We conduct a systematic analysis of 183 backdoor defenses introduced in papers published in top-tier AI and security venues and listed in Table~\ref{tab:papers_evaluated}. Our investigation uncovers serious methodological issues that undermine confidence in reported defense performance:

\begin{itemize}
    \item \textbf{Unrealistic threat models:} The vast majority of defenses are evaluated exclusively against outdated, dirty-label attacks (where attackers control training labels). Clean-label attacks, which preserve correct labels and represent more realistic threats, are rarely considered. Similarly, defenses are almost never tested against adaptive attackers who are aware of the defense mechanism.
    
    \item \textbf{Toy datasets dominate:} Over 73\% of defenses are evaluated on MNIST or CIFAR-10. ImageNet-1K, a more challenging benchmark, appears in only $\approx$20\% of papers. This raises serious concerns about generalizability to real-world deployment scenarios.
    
    \item \textbf{Limited trigger diversity:} Most evaluations use simple, visible patch-based triggers. Stealthy, imperceptible triggers are rarely considered, despite being more representative of sophisticated attacks.
    
    \item \textbf{Missing practical constraints:} Critical deployment considerations such as execution time, memory requirements, and computational cost are almost never reported, preventing practitioners from assessing real-world feasibility.
\end{itemize}

These evaluation gaps create a false sense of security: defenses appear highly effective on toy benchmarks but may fail under realistic threat models. When we re-evaluate selected defenses on more challenging datasets and against adaptive, stealthy attacks, we observe substantial performance degradation that contradicts published claims.

\textbf{Our contributions.} 
This paper presents the first systematization of knowledge (SoK) focused exclusively on backdoor defense evaluation methodology. Rather than proposing new defenses or attacks, we analyze \textit{how} the research community evaluates defenses and identify critical gaps that prevent meaningful comparison and deployment. Our work makes the following contributions:

\begin{itemize}
    \item \textbf{Comprehensive literature analysis:} We systematically survey 183 backdoor defense papers from top-tier venues (2018--2025), cataloging their evaluation methodologies, threat models, datasets, metrics, and experimental configurations.
    
    \item \textbf{Evaluation gap identification:} We identify critical methodological issues in defense evaluation, including the reliance on weak threat models, toy datasets, static triggers, and missing adaptive attacker evaluations.
    
    \item \textbf{Empirical evaluation:} We implement and evaluate 15 representative defenses across different categories (pre-training, in-training, post-training) against five diverse attacks (BadNets, Blended, WaNet, AdvDoor, Narcissus) on datasets of varying difficulty. Our results demonstrate that defense performance does not generalize from toy to realistic benchmarks. 
    
    \item \textbf{Actionable recommendations:} We provide concrete guidelines for defense researchers and practitioners, including minimum evaluation requirements (dataset difficulty, attack diversity, adaptive attacker consideration) and best practices for reproducible, fair comparison.

    \item \textbf{Public dataset:} Our comprehensive literature analysis is publicly available in our code repository as a dataset in a \texttt{.csv} file format, which is available for contributions from the community.
\end{itemize}

Our literature and experimental analysis aim to provide an objective and constructive examination of the current state of research on backdoor defenses. Rather than a critique of individual works, our goal is strictly to provide a clear, data-driven perspective on prevalent methodological trends and potential gaps within the literature, thereby fostering a shared commitment to enhancing the rigor and advancement of defense mechanisms. Moreover, the successful integration and reproduction of the representative defense methods selected for our experimental analysis speaks highly of their academic rigor and transparency.
\section{Background}
\label{sec:background}

\subsection{Backdoor Attacks}
Backdoor attacks are training-time attacks that add hidden behavior to the model~\cite{gu2019badnets}. The attacker's goal is two-fold: the first is to achieve high performance with clean (unaltered) samples, representing the clean behavior, and the second is to achieve high performance with malicious (modified) samples, representing the malicious behavior. Backdoor attacks can happen by poisoning the data~\cite{gu2019badnets}, injecting malicious code (code poisoning)~\cite{bagdasaryan2021blind}, or directly altering the model's weights (model poisoning)~\cite{hong2022handcrafted}. 

We briefly describe the relevant terminology of the backdoor attacks and their categorization; more information can be found in~\cite{li2022backdoor}.
There are various methods for injecting a backdoor into a model. These can be broadly divided into dirty and clean label backdoor attacks. First, the attacker chooses a trigger and adds it to a subset of the training data. Formally, let the training dataset be denoted as $\mathcal{D} = \{ ( \mathbf{x}_i, y_i )\}^N_{i=0}$, where $\mathbf{x}_i \in \mathcal{X}$ represents an input sample and $y_i \in \mathcal{Y}$ its corresponding ground truth label. The attacker's chosen trigger $t$ is added to a subset of training data. The proportion of poisoned data is given by $\epsilon = \frac{M}{N}$ where $M$ is the number of altered samples and $M \ll N$. By adding the trigger on each chosen sample, the attacker creates a malicious sample $\hat{\mathbf{x}} = \varphi(\mathbf{x}, t)$, where $\varphi$ is a trigger creation function. If the attacker follows a dirty label strategy, the ground truth label is changed to an attacker-chosen one $\hat{y}$~\cite{gu2019badnets}. For a clean label strategy, the ground truth label remains unchanged~\cite{turner2019label}. The malicious data set is constructed as $\hat{\mathcal{D}} = \{ ( \hat{\mathbf{x}}_i, \hat{y}_i )\}^M_{i=0} $ for dirty label attacks or $\hat{\mathcal{D}} = \{ ( \hat{\mathbf{x}}_i, y_i )\}^M_{i=0} $ for clean label attacks.

Depending on the trigger generation function $\varphi(\cdot)$, triggers are classified as static or dynamic. The static triggers, also known as sample agnostic, remain the same regardless of the input sample. Dynamic triggers, also known as sample-specific, vary depending on the input, i.e., they are unique per sample.
Depending on how we choose which samples to poison, we divide the attacks into source-agnostic and source-specific~\cite{guo2022overview}. In source-agnostic attacks, the trigger activates the backdoor only when applied to a sample of a specific class. On the other hand, in the source-agnostic attacks, the trigger can activate the backdoor in any sample.

The attack's performance is measured through attack success rate (ASR), which is the number of backdoor activations over the number of triggered inputs fed to a poisoned model. Clean accuracy, on the other hand, represents the behavior of the model on clean samples.

\subsection{Backdoor Defenses}
\label{sec:background-defenses}

\paragraph{\textbf{Deployment Stage}}
In the literature, backdoor defenses are categorized according to the stage of the machine learning pipeline at which they are applied~\cite{wu2022backdoorbench}. This classification distinguishes defenses based on the defender’s capabilities and the available resources. Specifically, we consider three main categories: 
\textit{pre-training}, \textit{in-training}, and \textit{post-training} defenses~\cite{wu2022backdoorbench}.
\textit{Pre-Training} defenses are applied before the model is trained. These defenses aim to prevent the injection of backdoors by sanitizing or augmenting the training data, selecting robust model architectures, or applying secure initialization techniques. The defender controls the dataset and training setup but has not yet started the actual training process.
\textit{In-Training} defenses control the training phase of the model. That is, the defender has access to the training data, the model, the weights, and the training hyperparameters. Furthermore, the defender controls the entire training procedure and can modify it to be robust against backdoors.
\textit{Post-Training} defenses analyze the model after it has been trained. These defenses assume that the defender does not participate in the training process but can analyze and modify the trained model.

\paragraph{\textbf{Domain}}
Defenses can be categorized into three domains: input-, feature-, and parameter-based, depending on where they are applied.
\textit{Input-space} defenses focus on the samples given to the model. Such defenses evaluate or check the model's input before passing it to the model.
\textit{Feature-space} defenses analyze the latent-space (at some point in the model), after the samples are passed to the model, to identify anomalies.
\textit{Parameter-space} defenses analyze the parameters of the model during training or after the training, and look for outliers or other indications that can be connected with a backdoor.

\paragraph{\textbf{Execution Stage}}
The defenses can also be categorized into two categories based on their deployment stage: \emph{Online} and \emph{Offline}~\cite{gao2020backdoor}.
\textit{Online} defenses work during the execution time and detect malicious samples or behaviors, on the fly. These types of defense are often bound by execution time.
\textit{Offline} defenses, on the contrary, are countermeasures that do not have a time constraint and are executed in a detached mode from the main execution process, for instance, checking the dataset before training, or analyzing the model's weights before deployment.

\paragraph{\textbf{Task}}
Backdoor defenses may perform several actions to prevent backdoor attacks. For example, one approach might involve identifying malicious instances within the training data, whereas another could determine whether or not a model contains a backdoor.
\textit{Malicious Sample Detection} defenses inspect the training dataset and look for malicious samples. The defenses could then further clean the dataset by removing the trigger from the malicious sample or just flagging that sample as malicious and removing it.
\textit{Backdoor Detection} defenses aim to detect the backdoor during training or after the training is completed.
\textit{Backdoor Removal} defenses remove the backdoor from the trained model after the backdoor is detected.


\subsection{Common Threat Models}
\label{sec:threat_models}

Backdoor attacks occur during the training phase, creating diverse threat scenarios depending on the attacker and defender capabilities. We categorize threat models along a spectrum of access and control, from full system access (white-box) to minimal interaction (black-box).\footnote{We provide some examples that do not represent the entire landscape of possible threat configurations.}

\paragraph{\textbf{White-Box}}

Both the defender and the attacker have full access to: (i)~model architecture and weights; (ii)~training algorithm, process, and hyperparameters; and (iii)~training data (the defender has no knowledge of whether the data is poisoned). For defenders, this scenario may occur when they use publicly available datasets from untrusted sources (e.g., Kaggle) to train their models from scratch or fine-tune pre-trained models from community repositories (e.g., Hugging Face). For attackers, this scenario is possible when, using their own data, they inject a backdoor in their own model and then release it for public use.

\paragraph{\textbf{Black-Box}}

This represents the most restrictive threat scenario, in which both the defender and the adversary have severely limited access and control. The defender can only interact with the model by providing inputs and observing the corresponding outputs, without any visibility into or ability to modify the model’s internal parameters or architecture. Similarly, the adversary can influence the model only indirectly. 
For example, by submitting malicious inputs through public APIs or by manipulating physical objects in the environment (e.g., placing stickers on stop signs) with the goal of having these altered instances captured and included in the training data.

\paragraph{\textbf{Gray-Box}}

Anything that is not \textit{white-box} or \textit{black-box} belongs in this scenario. It contains a wide range of partial knowledge and capability combinations. For example, the defender has access to the model weights but cannot access the training data. Alternatively, an attacker can inject malicious samples into a subset of the training data, but cannot control the model or training procedure.

Across all the threat models, we assume that the defender has access to a clean holdout dataset, and a trusted set of benign samples not used during training, which will be used to measure the clean accuracy and as a reference for detecting anomalies. This is a standard assumption in the literature. In this work, we evaluate various defenses each with its own assumptions. As a result, we do not follow a specific threat model in our experiments.

\section{Literature Study}
\label{sec:literature study}

The research on backdoor defenses has gained significant attention in recent years. Between 2018\footnote{We started our analysis from 2018, as the first backdoor attacks were BadNets~\cite{gu2019badnets} and Blended~\cite{chen2017targeted} both introduced in 2017, and the first countermeasures on this topic, i.e., Spectral Signatures~\cite{DBLP:conf/nips/Tran0M18} and Fine Pruning~\cite{liu2018fine} were published in 2018.} and 2025, the number of backdoor defense papers presented at AI and security conferences increased substantially. Very few defenses were proposed before 2020, but the numbers began to rise notably from 2021 onward. 
Although 2025 data are incomplete, current results already show continued interest. 

Clearly, the number of published papers\footnote{As we consider only nine conferences, the total number of backdoor defense papers is significantly higher.} motivates a more systematic evaluation of approaches given in the literature. Many novel and interesting works have been published in the last few years; however, there is no framework to help researchers evaluate the different aspects of each defense and perform a thorough robustness evaluation, resulting in incomplete reported results. In this work, our aim is to establish methodologies and best practices to examine backdoor defenses, helping researchers design better solutions and readers select appropriate defenses for each use case.

\subsection{Methodology}
\label{sec:lit-method}

We conducted an extensive literature review by evaluating all the papers on backdoor defenses published in selected $A^*$ venues\footnote{According to popular conference ranking systems, such as \url{www.conferenceranks.com}.} across both the security (i.e., S\&P, USENIX Security, NDSS, and CCS) and AI communities (i.e., ICLR, ICML, NeurIPS, CVPR, and AAAI) from 2018 to 2025. A total of 183 papers were manually analyzed, with key information extracted systematically from each study and saved in a \texttt{.csv} file.\footnote{We shared this file in the repository to allow further analysis and contributions from the community.}
The extracted information contains:
\begin{compactitem}
    \item \textbf{Identification information}: defense name, conference name, and publication year.
    \item \textbf{Evaluation information}: defense's code availability, the model architectures and the datasets (along with their domains, e.g., images, text, graph, sound, etc.) used in the experiments, the baseline defenses that were used for comparison, and the types of tested backdoor attacks. To gain a detailed view of the attacks against which the proposed defense is evaluated, we keep track of the different attack characteristics, whether an adaptive attacker was tested, the names of the attacks, the assumptions related to the attacker's knowledge of the defense, and the poisoning rates used. 
    \item \textbf{Evaluation metrics}: the metrics reported by the authors in their experiments, organized in two categories: attack performance with or without defense, and the attack's stealthiness. Finally, we report whether clean and no-attack performance were evaluated.
    \item \textbf{Defense properties}: information needed to characterize a defense as discussed in Section~\ref{sec:background-defenses}. More specifically, the defense's category, domain, and deployment stage. 
    \item \textbf{Defender's knowledge}: the scenario where the defense is evaluated (e.g., Federated Learning), the defender's knowledge about the model (architecture, hyperparameters, and weights), the backdoor's existence, the trigger (e.g., BadNets), the trigger's properties (e.g., trigger size or location), the trigger's distribution, and if the attack is targeted/untargeted. 
    \item \textbf{Access to resources}: whether the defender has access to poisoned or clean data, the model's weights, or the feature maps. Additionally, we check whether the defense requires additional labeled or unlabeled data beyond the training set, or extra models (poisoned or clean) beyond the one evaluated.
    \item \textbf{Defender's capability}: whether the defender can train surrogate models, alter the model's inputs or weights, fine-tune the model, and filter or purify malicious data.
\end{compactitem}
Observe that the last three categories are related to the defender's threat model.

\begin{table}[t]
\scriptsize
\centering
\caption{Number of analyzed papers per year and conference. ``-'' denotes no papers, and N/A indicates that the conference has not yet occurred. The table contains 181 papers. However, in our analysis, we included STRIP~\cite{gao2019strip} and Fine Pruning~\cite{liu2018fine} since we observed that, while they are not published in any of the considered conferences, they are the second and the third most popular baseline defenses.}
\label{tab:papers_evaluated_compressed}
\resizebox{\columnwidth}{!}{%
\begin{tabular}{lccccccccc}
\toprule
Year & S\&P & NDSS & CCS & USENIX & NeurIPS & ICML & ICLR & AAAI & CVPR \\ \midrule
2025 & 3 & 5 & 2 & 1 & N/A & 7 & 9 & 5 & 4 \\
2024 & 5 & 4 & 1 & 4 & 12 & 9 & 4 & 7 & 2 \\
2023 & 5 & 3 & 1 & 5 & 17 & 4 & 4 & 5 & 7 \\
2022 & - & 1 & - & 2 & 10 & 1 & 5 & 1 & 4 \\
2021 & - & 2 & - & 2 & 4 & 3 & 2 & 2 & - \\
2020 & - & - & - & - & - & 1 & 1 & - & 1 \\
2019 & 1 & - & 1 & - & 1 & - & - & - & - \\
2018 & - & - & - & - & 1 & - & - & - & - \\

\bottomrule
\end{tabular}
}
\end{table}


\subsection{Analysis of the Collected Data}

We present now relevant statistics from the collected information about existing defenses to support our discussion and experimental choices for this work, as well as to highlight current trends. We organize our findings into two categories: Recommendations (R), presented in light gray boxes, which are actionable suggestions with known solutions calling for widespread implementation and adoption; and Challenges (C), appearing in darker gray boxes, which are observed shortcomings requiring further research.

We note that $\approx80\%$ of the considered publications provide source code. Although this percentage is high, this still highlights the fact that many papers do not release their source code, while it greatly simplifies reproduction and validation, and enables extensive experimental analysis like the one we report in Section~\ref{sec:experiments}.

\begin{recommendation}
\label{rec:provide source code}
    Every published backdoor defense (and attack) paper should provide source code.
\end{recommendation}


Focusing on the deployment phase, we found that most defenses fall into the post-training category, accounting for nearly $60\%$ of all analyzed approaches. In-training defenses represent the second-largest group, accounting for $33\%$ of the total. In contrast, pre-training defenses and hybrid methods, such as those combining pre- or in-training with post-training techniques, are relatively rare, less than $5\%$. Finally, less than $1\%$ considers all stages of the training process.
These different categories assume different attack and defense threat models and capabilities. Post-training is commonly considered a worst-case scenario for defenders as it is a reactive stage (end-user) where defense assumes the attack has already succeeded. Thus, the trend of focusing on a reactive defender neglects more effective stages for defending against backdoor attacks (pre- and in-training). 

\begin{recommendation}
\label{rec:different defense stages}
    Balance the study of different defense deployment phases by focusing on proactive defense methods.
\end{recommendation}

Moreover, the diverse settings of existing backdoor attacks require defenses that are flexible in different deployment scenarios, and limiting a defense to operate in a single phase in most cases requires assumptions that may be unrealistic in certain scenarios. Nowadays, we rely heavily on pre-trained models for transfer learning, which may have been trained on malicious data inaccessible to the user downloading them, making pre-training defenses impractical. At the same time, when we have access only to data and must rely on online services for ``black-box'' training or fine-tuning, due to limited resources or the use of closed models (e.g., OpenAI\footnote{\url{https://openai.com/index/gpt-4o-fine-tuning/}}), we cannot control or inspect the resulting model and can only use it as-is. This emphasizes the need for defenses that remain effective and adaptable across different phases.


\begin{figure}[!ht]
    \centering
    \includegraphics[width=\linewidth, trim=20pt 22pt 0pt 0pt, clip]{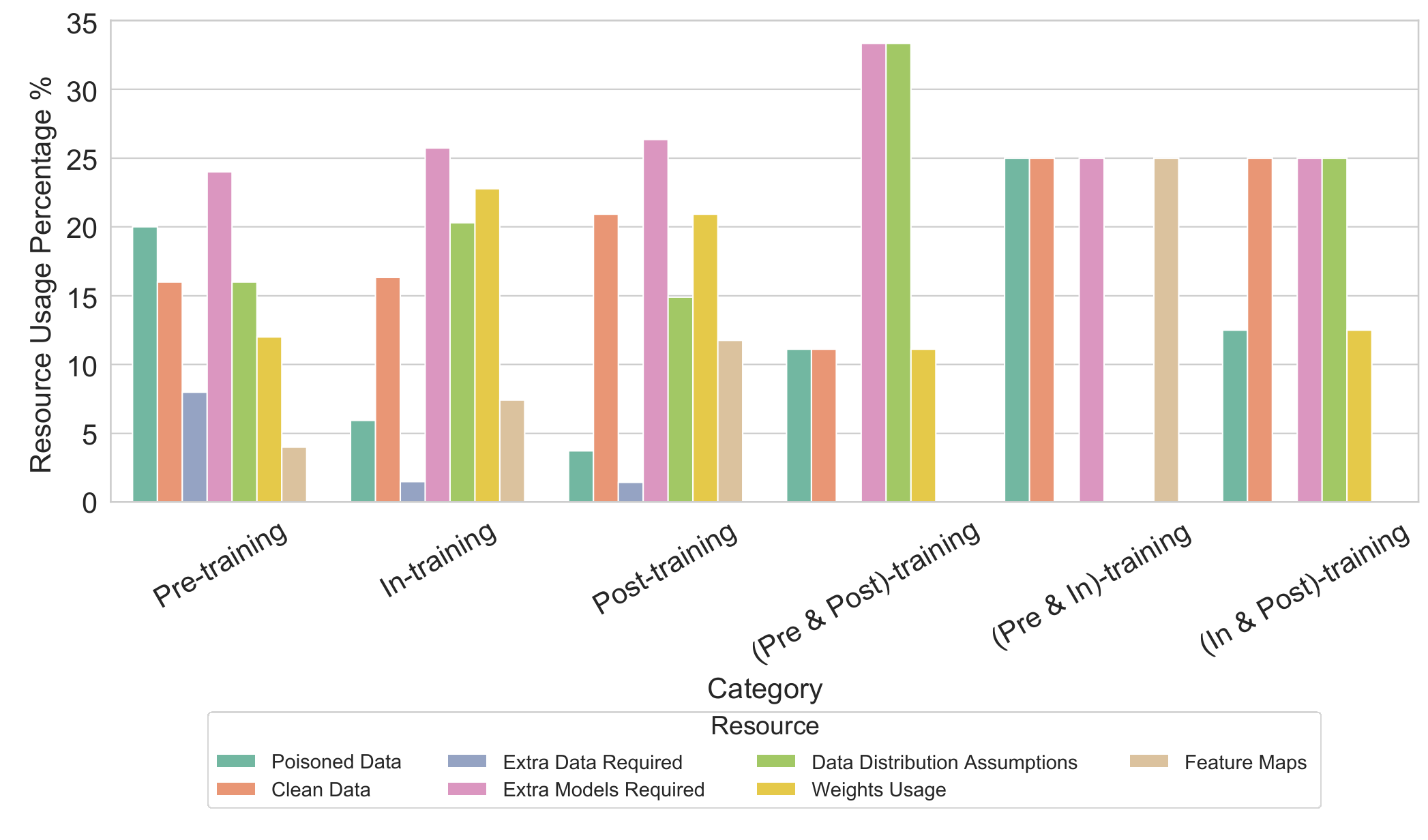}
    \caption{Distribution of resources across different training phases, showing the relative occurrence of each resource type during pre-, in-, and post-training phases.}
    \label{fig:category_distribution_resources}
\end{figure}

If we consider defenses grouped by categories as shown in Figure~\ref{fig:category_distribution_resources}, we can see how defenses in each category use different resources. A common requirement is having auxiliary models, which could not be realistic due to computational constraints when dealing with large models or datasets, since they have to be trained, often from scratch. This aspect is further discussed in Section~\ref{sec:experiments} where we show how this could limit the practical use of these defenses.
The second most used resource for pre-training defenses is the reliance on poisoned data. This constitutes a significant assumption in practical scenarios because it requires access to every potential attack to ensure resilience, thereby restricting the generalizability of defenses. 

Finally, many methods assume access to clean training data, an unrealistic assumption in real-world settings where models are distributed to large user communities, as seen with the HuggingFace repositories. In such cases, a more realistic assumption would imply access to data exhibiting the same feature distribution as the training set, without requiring the exact data points used during training.

\begin{challenge}
\label{chall: resource related challenge}
    Develop scalable defenses independent of auxiliary models and poisoned/clean training data, using alternative resources  (e.g., feature maps or same-distribution samples not from the training set) to improve generalization and real-world performance.
\end{challenge}

\begin{figure}[!ht]
    \centering

    \begin{subfigure}[b]{0.49\linewidth}
        \centering
        \includegraphics[width=\linewidth]{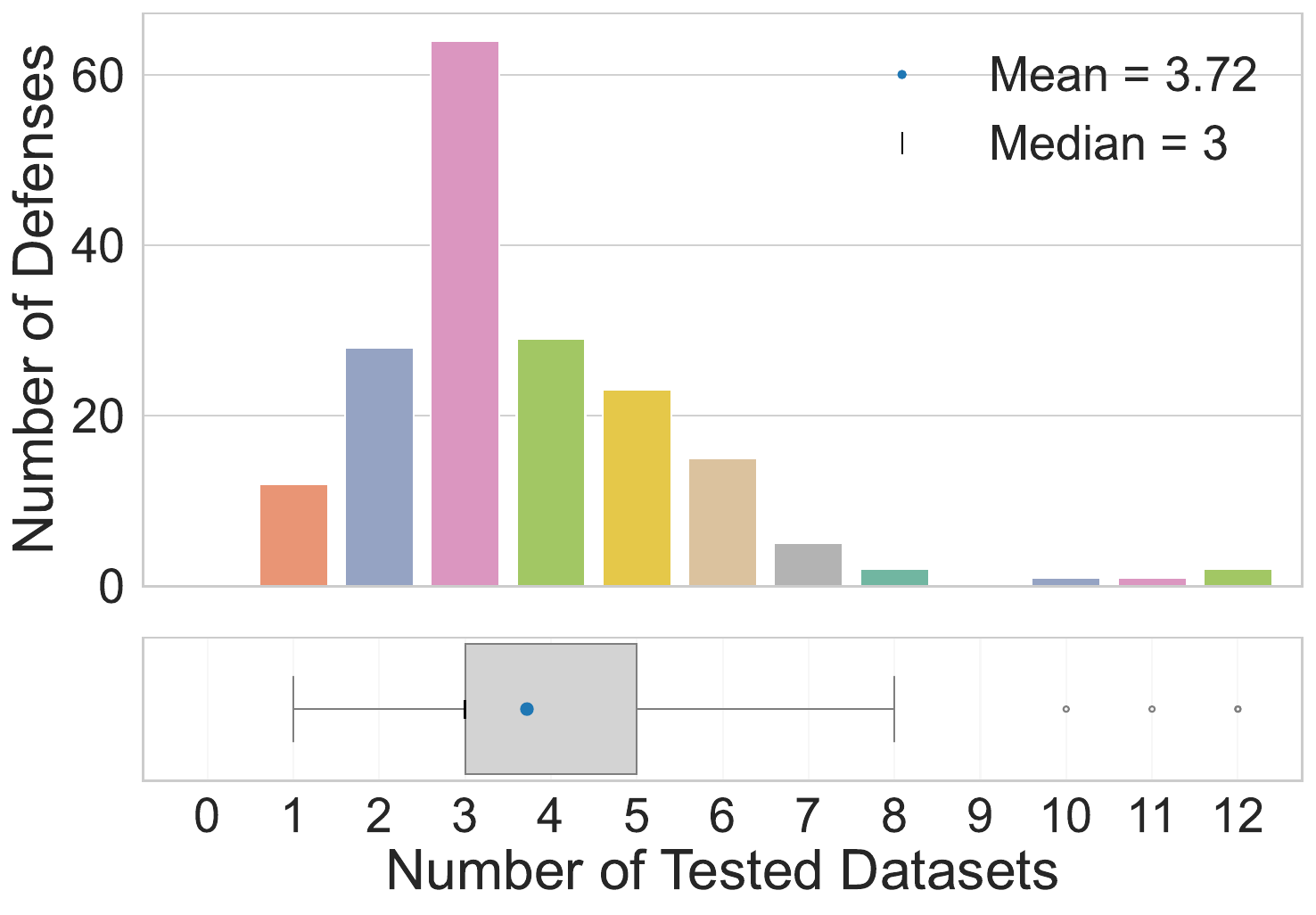}
        \caption{Datasets}\label{fig:tested-datasets}
    \end{subfigure}
    \begin{subfigure}[b]{0.49\linewidth}
        \centering
        \includegraphics[width=\linewidth]{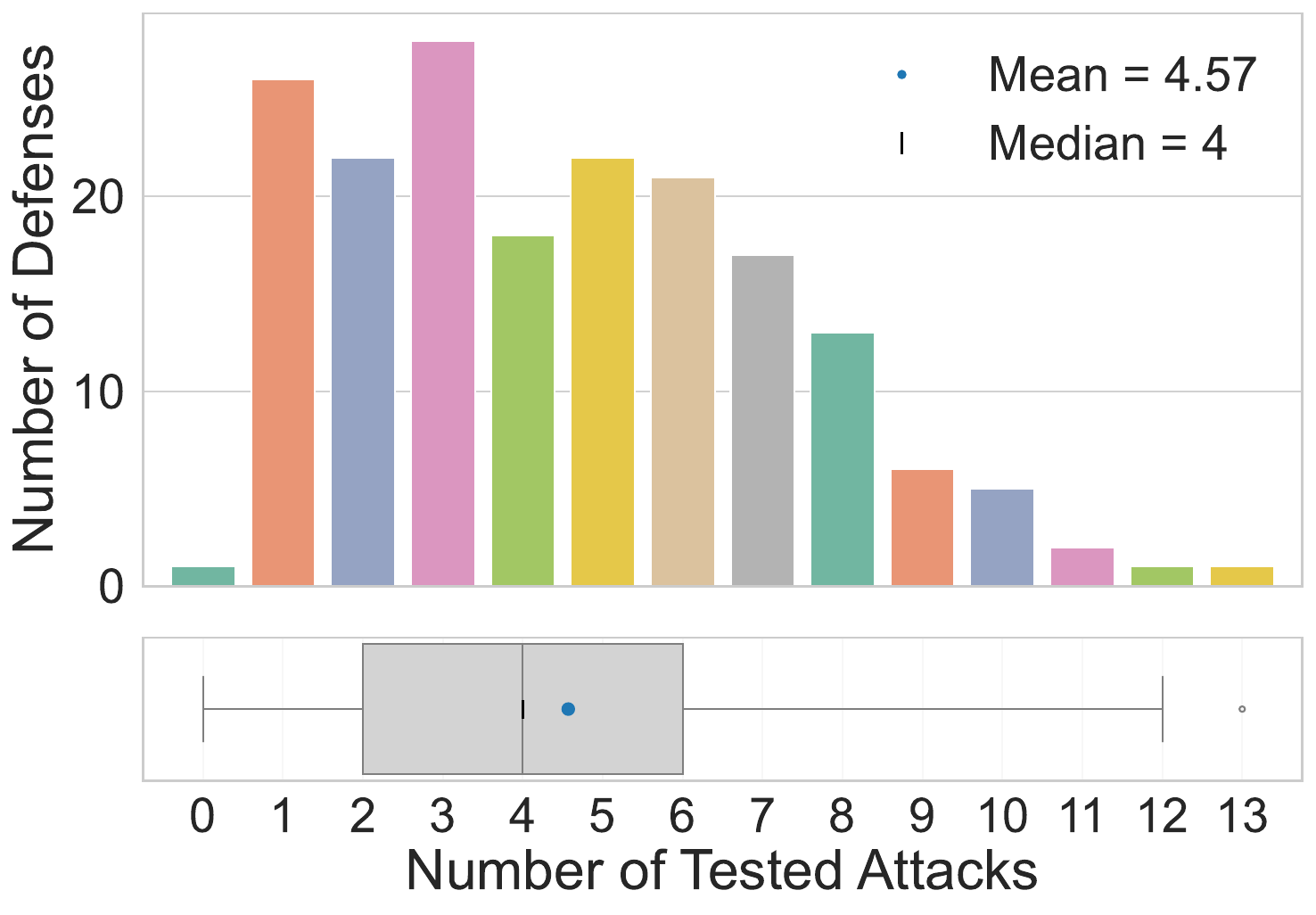}
        \caption{Attacks}\label{fig:tested-attacks}
    \end{subfigure}
    \\
    \begin{subfigure}[b]{0.49\linewidth}
        \centering
        \includegraphics[width=\linewidth]{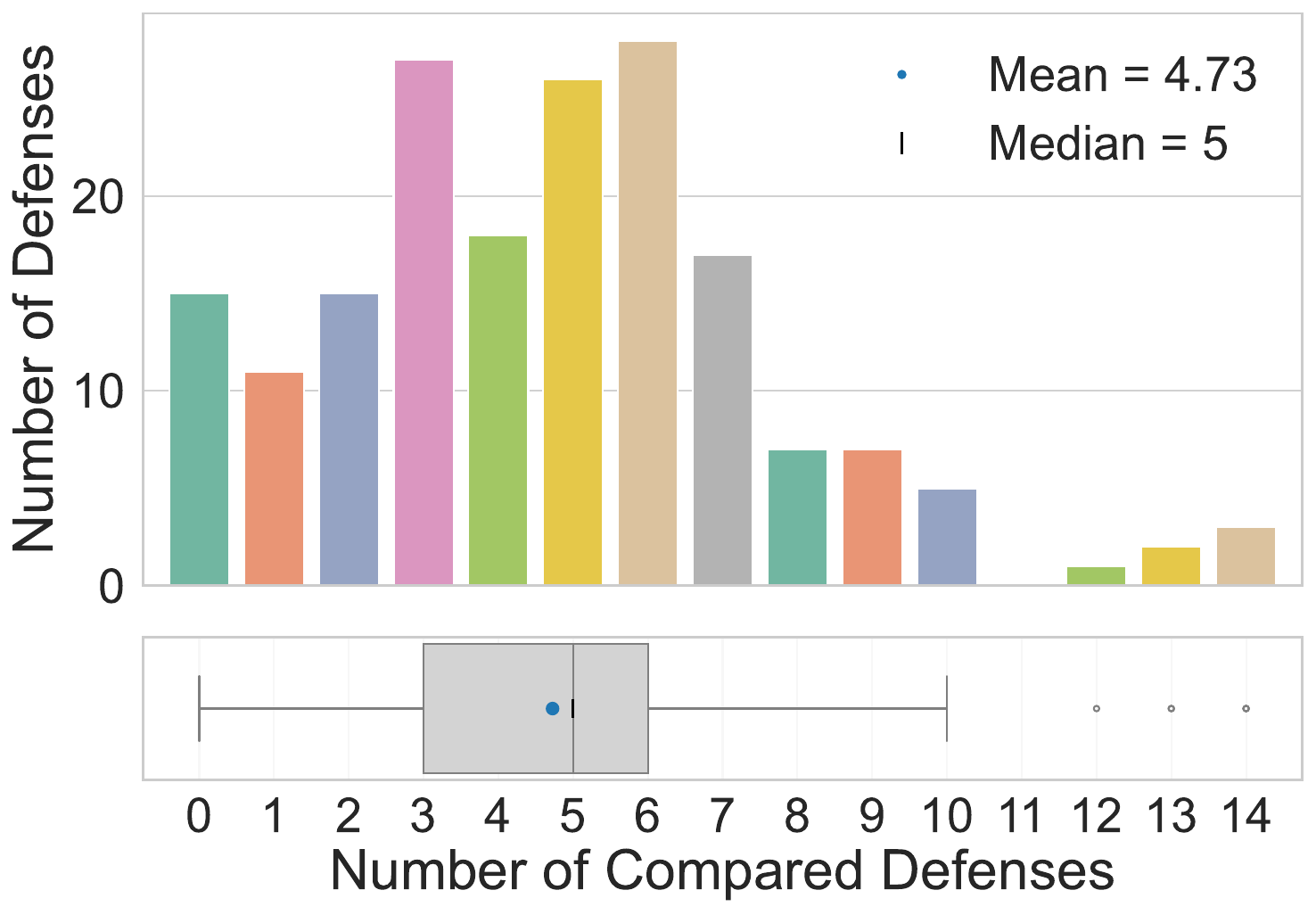}
        \caption{Defenses}\label{fig:tested-defenses}
    \end{subfigure}
    \begin{subfigure}[b]{0.49\linewidth}
        \centering
        \includegraphics[width=\linewidth]{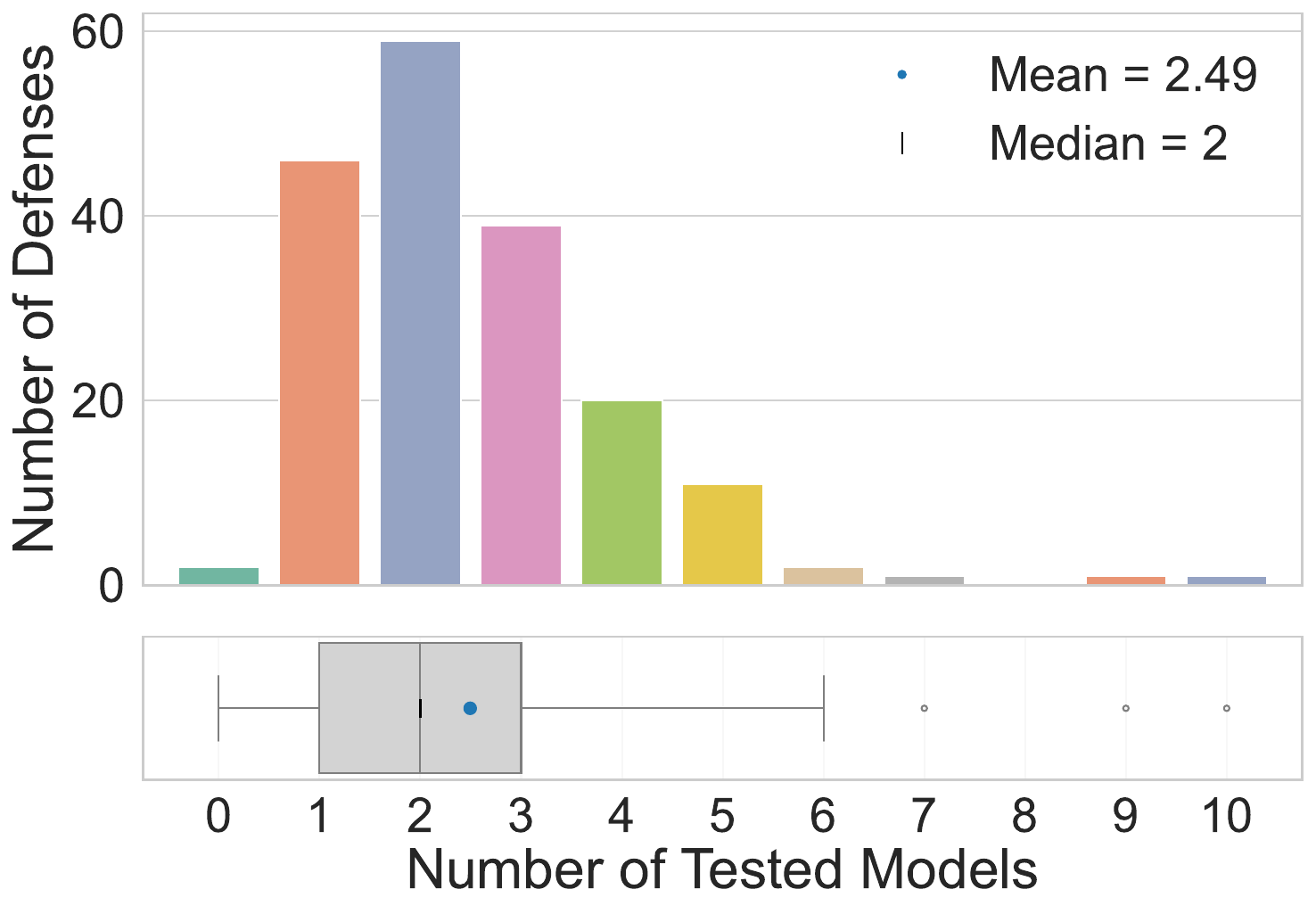}
        \caption{Model types}\label{fig:tested-models}
    \end{subfigure}

    \caption{Distribution of the number of used datasets, attacks, defenses, and model types used in the analyzed defenses.}
    \label{fig:top5_datasets_per_conference_type_and_year}
\end{figure}

In Figure~\ref{fig:top5_datasets_per_conference_type_and_year}, we provide statistics on the number of datasets, attacks, types of models, and baseline defenses used to evaluate performance. As shown in Figure~\ref{fig:tested-datasets}, most defenses are tested using three datasets, while considering more than six datasets is rare. Similarly, in most cases, one to three different model types are used (see Figure~\ref{fig:tested-models})\footnote{See Appendix~\ref{sec:model type groups} for the utilized model groups for this analysis.}. The performance comparison is done against five or six baseline defenses,\footnote{Notably, in our sample, some papers do not even compare against any defense~\cite{cao2020fltrust,DBLP:conf/nips/0002F23,DBLP:conf/iclr/DuJS20,DBLP:conf/aaai/DoanLY023}. In one case also, the proposed theoretical framework against data poisoning was not tested against any specific attack, but only against distribution shift~\cite{DBLP:conf/nips/0002F23}.}
and the number of targeted attacks is typically between one and five.

Finally, we observed that, overall, only about $58\%$ of the proposed defenses consider an adaptive attacker, around $55\%$ of the defenses presented at AI-focused conferences, and roughly $67\%$ of those presented at security conferences. While considering this aspect is crucial, given that adaptive attacks are generally the most advanced, it alone does not suffice to demonstrate the effectiveness of the defense. 
Overall, this highlights the heterogeneity of evaluation settings, which limits the comparability of results across studies and underscores the need for more comprehensive and standardized benchmarking.

\begin{figure}[!ht]
    \centering
    \subfloat[Top 10 most used datasets.\label{fig:top_datasets_ai_vs_security}]{
        \includegraphics[width=0.48\linewidth, trim= 5pt 11pt 7pt 6pt, clip]{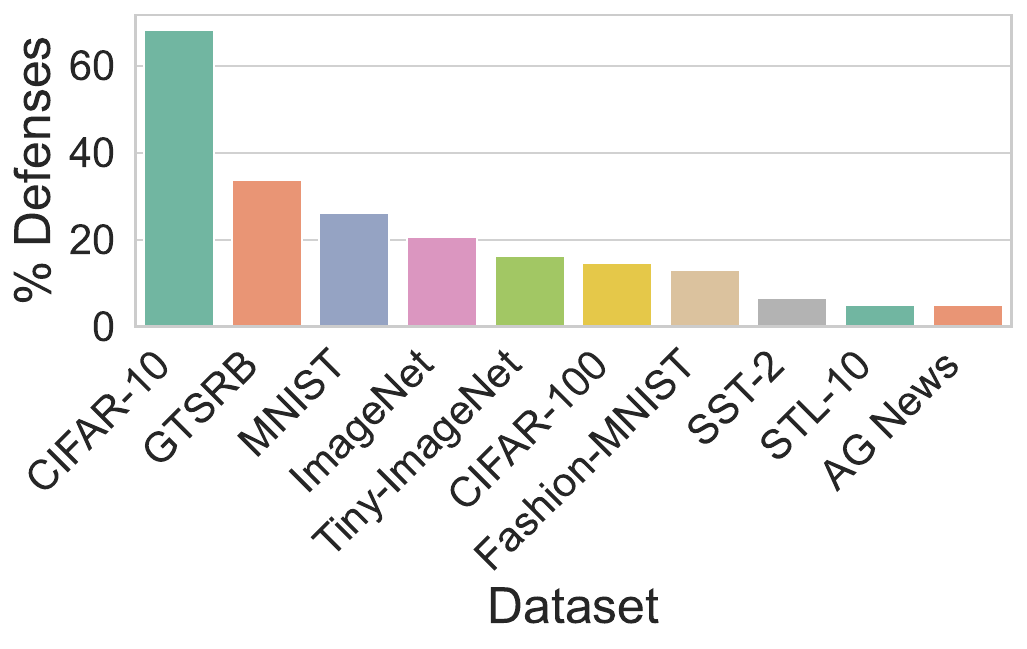}
    }
    \subfloat[Top 10 most used models.\label{fig:models_used_per_year}]{
        \includegraphics[width=0.48\linewidth, trim= 5pt -25pt 7pt 6pt, clip]{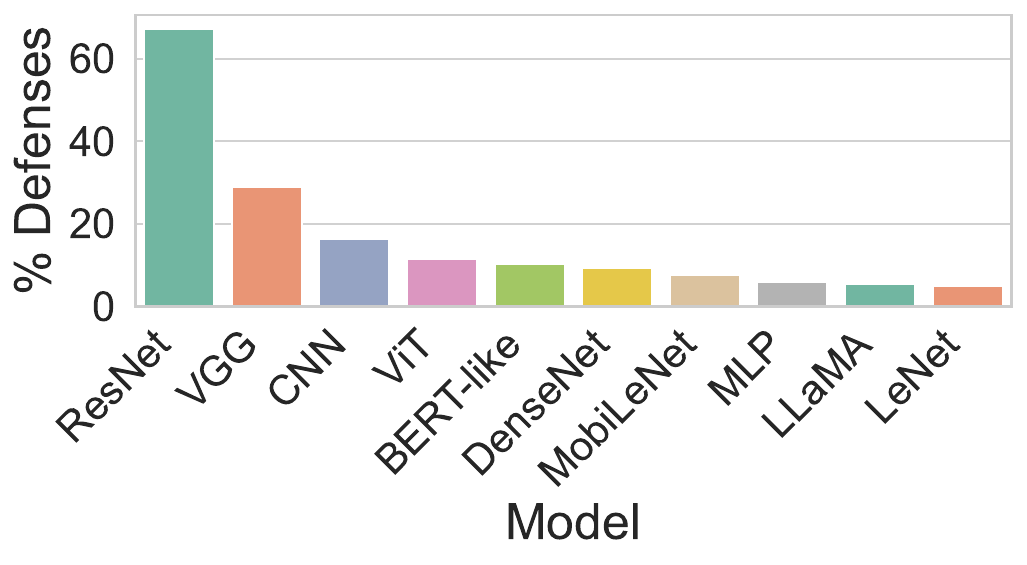}
    }
    
    \vspace{1em}
    
    \subfloat[Top 5 most evaluated defenses.\label{fig:top5_defenses_conference_type_percentage}]{
        \includegraphics[width=0.48\linewidth, trim=4pt 9pt 7pt 7pt, clip]{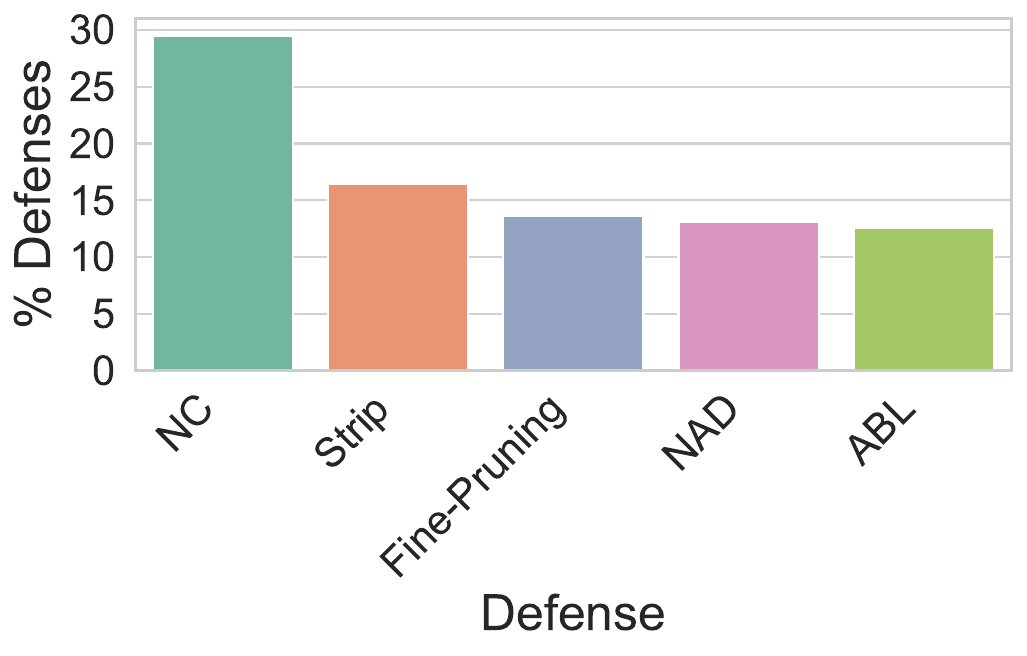}
    }
    \subfloat[Top 5 most tested attacks.\label{fig:top5_attacks_conference_type_percentage}]{
        \includegraphics[width=0.48\linewidth, trim=0pt -7pt 0pt -7pt, clip]{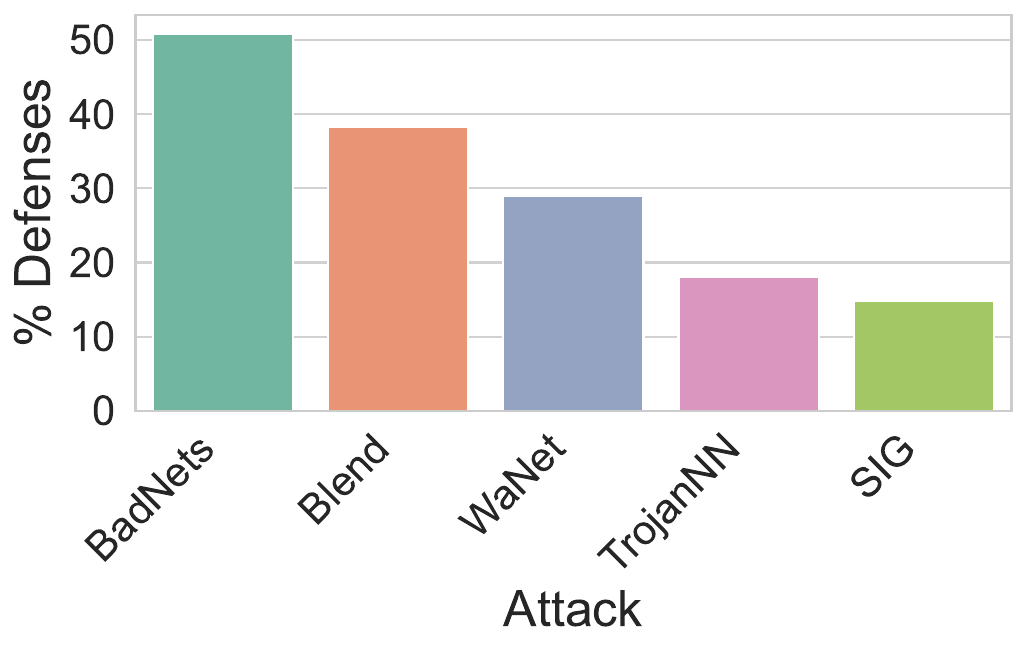}
    }

    \caption{Overview of dataset, attack, defense, and model evaluation statistics.}
    \label{fig:four_stats}
\end{figure}

Another important trend visible in Figure~\ref{fig:top_datasets_ai_vs_security} is the strong tendency to consider a narrow subset of available benchmark datasets. In particular, CIFAR-10 emerges as the predominant choice, serving as the baseline for most evaluations. While this facilitates comparability across studies, it also raises important concerns about the general applicability of the evaluation results for the proposed defenses. 

Figure~\ref{fig:top5_attacks_conference_type_percentage} shows the most common attacks used in the experimental evaluations of the proposed defenses. Among these, Badnets and Blend attacks are the most common ones. The trend highlights a continued reliance on a narrow and well-established (though arguably outdated) set of evaluation techniques.
Despite the rapid growth of the literature on adversarial machine learning, only a small fraction of defenses evaluate their approaches on recent attacks. Among the five most frequently considered attacks, the most recent is SIG~\cite{fields2021trojan}, introduced in 2021, thus highlighting a gap between the development of new attacks and their adoption in defense evaluations. 
Therefore, while defenses may appear robust under current experimental standards, their true resilience against state-of-the-art or emerging attacks remains underexplored. Most attacks discussed in Section~\ref{sec:lit-method} are dirty-label, with clean-label attacks largely overlooked. Additionally, triggers are typically visible rather than stealthy, and only one case involved a dynamically generated trigger, highlighting limited attention to complex or adaptive designs. The same observation also holds for the baseline defenses most frequently used for comparison with the proposed methods. Notably, the most recent of these is ABL~\cite{DBLP:conf/nips/LiLKLLM21} from 2021, also in this case.

In Figure~\ref{fig:models_used_per_year}, we examine the machine learning models used for the evaluation of the analyzed defenses.\footnote{See Appendix~\ref{sec:model type groups} for the model groups.} Typically, used models belong to the ResNet group that includes different variants of ResNet models, such as ResNet18 or ResNet50. This raises the concern that defenses may be overly tuned to a specific architecture, limiting their adaptability across different model families. Because of this, defenses might potentially exploit unique characteristics of specific target models, such as residual blocks or simple CNNs layers, while neglecting a wider set of characteristics from diverse model architectures. As a consequence, this can effectively impose an expiration date on the defense’s applicability.
Additionally, we observe a preference for simpler models, which may require less time and resources to train.
From Figure~\ref{fig:four_stats}, we can conclude that focusing only on a limited selection of toy datasets, specific model families, or certain attacks and defenses can lead to defenses that are overfitted to these particular scenarios, limiting their applicability beyond these predefined assumptions.

\begin{challenge}
    Design defenses that operate at a higher level of abstraction, avoiding assumptions tied to specific architectures, such as CNNs, or data, to ensure future-proof robustness across models and domains. 
\end{challenge}

Thus, the number of evaluated attacks/defenses, models, and datasets should be less important than ensuring that diverse categories are covered. This implies using both simple datasets (e.g., MNIST) and complex ones (e.g., ImageNet-1K), as well as simpler model architectures (e.g., VGG family) and more advanced ones (e.g., ViT), as proved by the results we collected in Section~\ref{sec:experiments}. Finally, for attacks and defenses, it is important to use both well-established baselines and recent state-of-the-art results. 

\begin{recommendation}
\label{rec: diversity in evaluation}
    Prioritize diversity in the evaluation setup by including datasets of varying complexity, models from different families, and both established and recent attacks and defenses, rather than simply increasing their number.
\end{recommendation}

Next, we examine the distribution of defenses based on the domain of the datasets considered in the paper. We observed that the image domain is the most frequently used, with the text domain receiving more attention starting from 2022. Other domains are not as prominent, although in recent years, we can observe a greater diversity in the dataset domain than in earlier years. This variation could be attributed to the rise of Large Language Models (LLMs), though the prevalence of the image domain still remains. As LLMs continue to advance and evolve into multi-modal systems such as visual language models (VLMs), which integrate both textual and visual inputs, the boundaries between input domains are becoming increasingly blurred, making domain-specific defenses less appealing.

\begin{challenge}
     Design defenses that are dataset domain-agnostic.
\end{challenge}


\begin{table}[h!]
\centering
\caption{Percentage of defenses per defense domain that can work over a Single or Multiple dataset domains (absolute count in parentheses).}
\label{tab:domain_defense_normalized_transposed_counts_clean}
\begin{tabular}{lccc}
\hline
\textbf{Defense Domain} & \textbf{Single (count: 140)} & \textbf{Multiple (count: 20)} \\
\hline
Input-Space     & 48.6\% & 20.0\% \\
Feature-Space   & 23.6\% & 15.0\% \\
Parameter-Space & 44.3\% & 70.0\% \\
\hline
\end{tabular}
\end{table}

We then analyze the relationship between the action space of the defenses (Input, Feature, or Parameter) and the dataset domain (e.g., images or text) in Table~\ref{tab:domain_defense_normalized_transposed_counts_clean}.
Defenses that operate on model parameters are more frequently tested against a wider range of dataset domains, such as both image and text. This suggests that access to model parameters allows for the development of defenses that are more broadly applicable across different types of data. 
While designing defenses that operate across domains is valuable, overly general approaches can reduce effectiveness. In many industrial settings, more specialized defenses, tailored to a specific application, often achieve higher reliability. Therefore, a careful trade-off between generality and specificity should be considered when developing new defenses.

\begin{recommendation}
    Combine general, model-agnostic defenses with domain and pipeline-specific methods to achieve transferability.
\end{recommendation}

Figure~\ref{fig:performance_metrics_by_conference_type_percentage} in Appendix~\ref{app:metrics} depicts the usage patterns of evaluation metrics across defenses. 
We can broadly categorize metrics into three groups: \textbf{(i) task performance:} Clean Accuracy (CA), ASR, and Certified Accuracy; \textbf{(ii) detection capability:} Detection Accuracy, Precision, Recall (TPR), False Positive Rate (FPR), F1-Score, and AUC-ROC; \textbf{(iii) efficiency:} Execution Time. 
From our analysis, we found that most adopted metrics are ASR, followed by CA in AI conferences, and a set of Other different metrics and CA at security conferences. The difference is quite large, with $\approx65\%$ and $\approx80\%$ of defenses reporting ASR, but only $\approx48\%$ and $\approx39\%$ report CA at AI and Security conferences, respectively. 
The absence of a standardized and consolidated set of evaluation metrics limits the comprehensive comparison of proposed defenses. Although ASR serves as an indicator of a defense’s capacity to block attacks, its true effectiveness also depends on the ability to maintain the model’s main task performance (commonly measured with CA) and ensure seamless deployment.
Moreover, defense evaluation papers rarely report execution time, despite its critical importance for practical deployment.

Stealthiness is a crucial property of a backdoor attack, tightly connected to the attack's performance against a defense. 
An attack could be stealthy in the input-, feature-, or parameter-space, and many defenses are not effective in all spaces simultaneously~\cite{xu2025towards}. 
Thus, to verify a defense's generalization, it is important to consider attacks that are stealthy in different spaces. The evaluator could identify stealthy attacks using various stealthiness metrics or pick attacks that are known to be stealthy in different spaces. For example, Adap-blend is stealthy in the feature space but not in the input space, WaNet is stealthy in the input space but not in the feature space, and Grond is stealthy in all spaces~\cite{xu2025towards}. Surprisingly, only 9 out of 183 surveyed papers report results about the attack's stealthiness, which shows that this topic is often overlooked in a defense evaluation. 


\begin{recommendation}
    When evaluating a defense, we need to consider attacks with different stealthiness capabilities (pixel space, latent space, parameter space, frequency domain), often measured by appropriate metrics, instead of testing many attacks with the same properties.
\end{recommendation}

Defenses must report execution time, a critical metric for practical deployment. Some theoretically effective defenses may fail in real-world applications due to high computational costs, yet this limitation remains unclear without systematic runtime reporting and comparison. Following recommendation~\ref{rec:report execution time} enables fair comparison of computational costs and helps practitioners assess deployment feasibility.

\begin{recommendation}
\label{rec:report execution time}
Report the defense's execution time and compare it against the baseline (no-defense) execution time, quantifying the induced overhead under identical hardware settings. 
\end{recommendation}

Depending on the nature of the defense, e.g., backdoor detection (model or input-level) or backdoor removal, different metrics will be reported as applicable. However, while we observe some commonly used metrics, there are still a number of papers reporting ``other'' metrics, which include less used ones, such as ``anomaly index''. These can be valid, as defense might be further from the already established types of defenses, but we might want to avoid arbitrary metrics. While core metrics (accuracy and ASR) are widely adopted, inconsistent naming conventions and the introduction of non-standard metrics without clear justification complicate reproducibility and cross-study comparisons.

\begin{challenge}
    Lack of standardized terminology and consistent reporting of metrics when evaluating backdoor defenses. 
\end{challenge}

\begin{figure}[!ht]
        \centering
        \includegraphics[width=\linewidth, trim=0pt 26pt 0pt 0pt, clip]{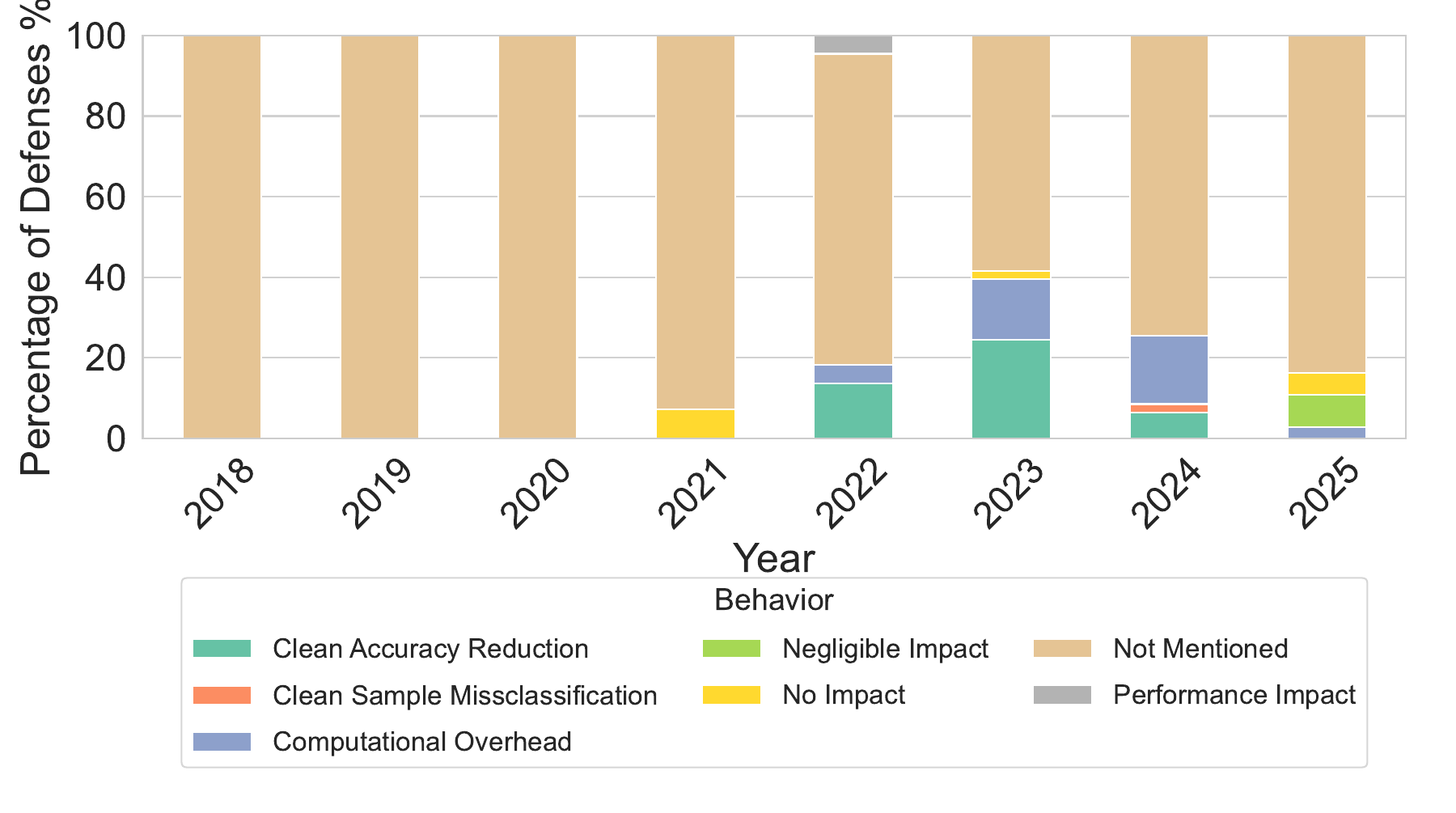}
        \caption{Percentage per year of defenses that consider the effect of the defense when there is no attack.}
        \label{fig:no attack behavior}
\end{figure}

Next, Figure~\ref{fig:no attack behavior} provides statistics (based on what the authors reported) on what happens once the defense is deployed on clean models. 
We can observe that in most cases, the effect is not discussed in the paper. Over the years, more papers have started to consider this, reporting (vague) information on it. 

\begin{recommendation}
\label{rec: report defense performance in no-attack case}
     Report the impact of the defense without any backdoor present, explicitly including the main task (clean) accuracy when the defense is applied.
\end{recommendation}


\section{Experimental Setup}

\paragraph{\textbf{Datasets}}
We evaluate on MNIST, CIFAR-100, and ImageNet-1K, selected for their diversity and difficulty (as in Recommendation~\ref{rec: diversity in evaluation}). MNIST is grayscale and simple; its inclusion ensures that grayscale images are considered in the evaluation, as such data frequently occurs in applications like medical imaging. CIFAR-100 offers mid-level complexity with 100 classes. ImageNet-1K provides a large-scale, realistic benchmark with 1M+ images across 1\,000 classes.

\paragraph{\textbf{Models}}
As seen in Figure~\ref{fig:four_stats}, the most used models are ResNet, VGG, and the ViT family, followed by custom CNN architectures and DenseNet. Thus, in our experiments, we consider ResNet-18~\cite{he2016deep}, VGG-19~\cite{simonyan2014very}, ViT-B/16~\cite{dosovitskiy2020image}, and DenseNet-121~\cite{huang2017densely}. These cover both convolutional and transformer-based architectures.

\paragraph{\textbf{Defenses}}
Defenses are selected based on \textbf{(i) code availability:} we include only defenses with publicly available, runnable implementations to ensure reproducible evaluation and avoid errors; \textbf{(ii) category coverage:} we select at least one defense from each major category combination (see Section~\ref{sec:background-defenses}) to provide comprehensive coverage of defense strategies; \textbf{(iii) recency:} when multiple defenses occupy the same category, we favor more recent work (2022-2024) to reflect the current state-of-the-art, unless older seminal work represents a fundamentally different approach; \textbf{(iv) integration complexity:} we prioritize defenses with standard interfaces and minimal implementation dependencies to enable fair comparisons. See Table~\ref{tab:defenses_for_experiments} for a complete list of the defenses considered for our experimentation.

\begin{table}[htbp]
\centering
\caption{Selected defenses categorized by training stage, inspection domain, and execution mode.}
\label{tab:defenses_for_experiments}
\resizebox{\columnwidth}{!}{%
\begin{tabular}{@{}p{2.5cm}p{2.5cm}p{2.2cm}p{4.5cm}@{}}
\toprule
\textbf{Deployment Stage} & \textbf{Inspection Domain} & \textbf{Execution Mode} & \textbf{Defense} \\
\midrule
\multirow{6}{*}{In-training} 
  & Feature & Offline & Spectral Signatures~\cite{DBLP:conf/nips/Tran0M18} \\
  & Feature & Online & Trap and Replace~\cite{DBLP:conf/nips/WangHZZW22} \\
  & Input & Online & Reback (Need for Speed)~\cite{DBLP:conf/sp/MaYLYLLQ24} \\
  & Input, Parameter & Online & PDB~\cite{DBLP:conf/nips/WeiZW24} \\
  & Parameter & Offline & ABL~\cite{DBLP:conf/nips/LiLKLLM21} \\
  & Parameter & Online & CBD~\cite{zhang2023backdoor} \\
\midrule
\multirow{7}{*}{Post-training} 
  & Feature & Offline & NAD~\cite{DBLP:conf/iclr/LiLKLLM21} \\
  & Feature, Input & Offline & Unicorn~\cite{DBLP:conf/iclr/WangMZM23} \\
  & Input & Offline & Neural Cleanse~\cite{DBLP:conf/sp/WangYSLVZZ19} \\
  & Input & Online & Scale-Up~\cite{DBLP:conf/iclr/GuoLCG0023} \\
  & Input, Parameter & Offline & AIBD~\cite{DBLP:conf/aaai/YinWLLL25} \\
  & Parameter & Offline & IBAU, BAN~\cite{xu2024ban} \\
  & Parameter & Online & BadExpert~\cite{DBLP:conf/iclr/XieQHLWM24} \\
\midrule
\multirow{2}{*}{Pre-training} 
  & Input & Offline & Incompatibility Clustering~\cite{DBLP:conf/iclr/JinSR23} \\
  & Input, Parameter & Offline & Metasift~\cite{DBLP:conf/uss/ZengPJ0LJ23} \\
\bottomrule
\end{tabular}
}
\end{table}



\paragraph{\textbf{Backdoor Attacks}}
\label{sec:attacks}

We evaluate defenses against five representative attacks (Table~\ref{tab:attack_summary}): BadNets~\cite{gu2019badnets}, Blended~\cite{chen2017targeted}, WaNet~\cite{nguyen2021wanet}, AdvDoor~\cite{zhang2021advdoor}, and Narcissus~\cite{zeng2023narcissus}, covering both input- and feature-space manipulations, and dirty- vs. clean-label settings.

\paragraph{\textbf{Poison rates}}
We selected four specific poison rates, 0.005 (0.5\%), 0.01 (1\%), 0.05 (5\%), and 0.2 (20\%), to comprehensively evaluate selected defenses. The rates of 1\% and 5\% were included because our analysis shows they are the most commonly used poisoning levels in the existing literature. We then add two edge-case rates: 0.5\%, representing a subtle attack used in a minority of defense papers (5\%), and 20\%, representing a less subtle attack (used in 10\% of defense papers).
We evaluated Narcissus with three poison rates (other attacks used five), because Narcissus is a clean-label attack. Matching the total number of poisoned images requires increasing the per-class poison rate, which is limited by class size (e.g., poisoning an entire class in CIFAR-100 equals only $1\%$ of the dataset). In the case of Imagenet-1K, we could not achieve a poison rate matching the baselines, even when poisoning an entire class, due to the large number of classes.
We also test poison rate 0, indicating a no-attack scenario, to evaluate the impact of the deployed defenses when the attack is not present. 

\section{Experimental Results}
\label{sec:experiments}

Excluding runs with timeouts and Unicorn and Neural Cleanse defenses, the 2\,798 successful experiments across all three datasets collectively required more than 170 days of total compute time.
All results are available in the repository\footnote{We will share our code and additional results after acceptance of the paper.}, while here we report the key results. The missing values in the presented results mean that the given defense failed in that specific configuration.
Table~\ref{tab:avg_results_raw} in Appendix~\ref{sec:additional_results} presents the average results of the attacks and models across the various datasets without defenses to facilitate comparison in the following case studies.


\paragraph{\textbf{\textit{Model Architectures}}}
\label{sec:CaseStudyArchitecture}


Based on the survey in Section~\ref{sec:literature study}, most existing defenses are developed for image-based attacks on CNN architectures. Transformer-based models, such as Vision Transformers (ViTs), which underpin many state-of-the-art generative models, remain largely unexplored. Given their growing importance, ViTs should serve as key baselines for evaluating backdoor defenses, highlighting the need for architecture-agnostic defenses.

\begin{figure}[!ht]
    \centering
    \begin{subfigure}[b]{0.8\columnwidth}
        \centering
        \includegraphics[width=\columnwidth, trim=10pt 16pt 10pt 10pt, clip]{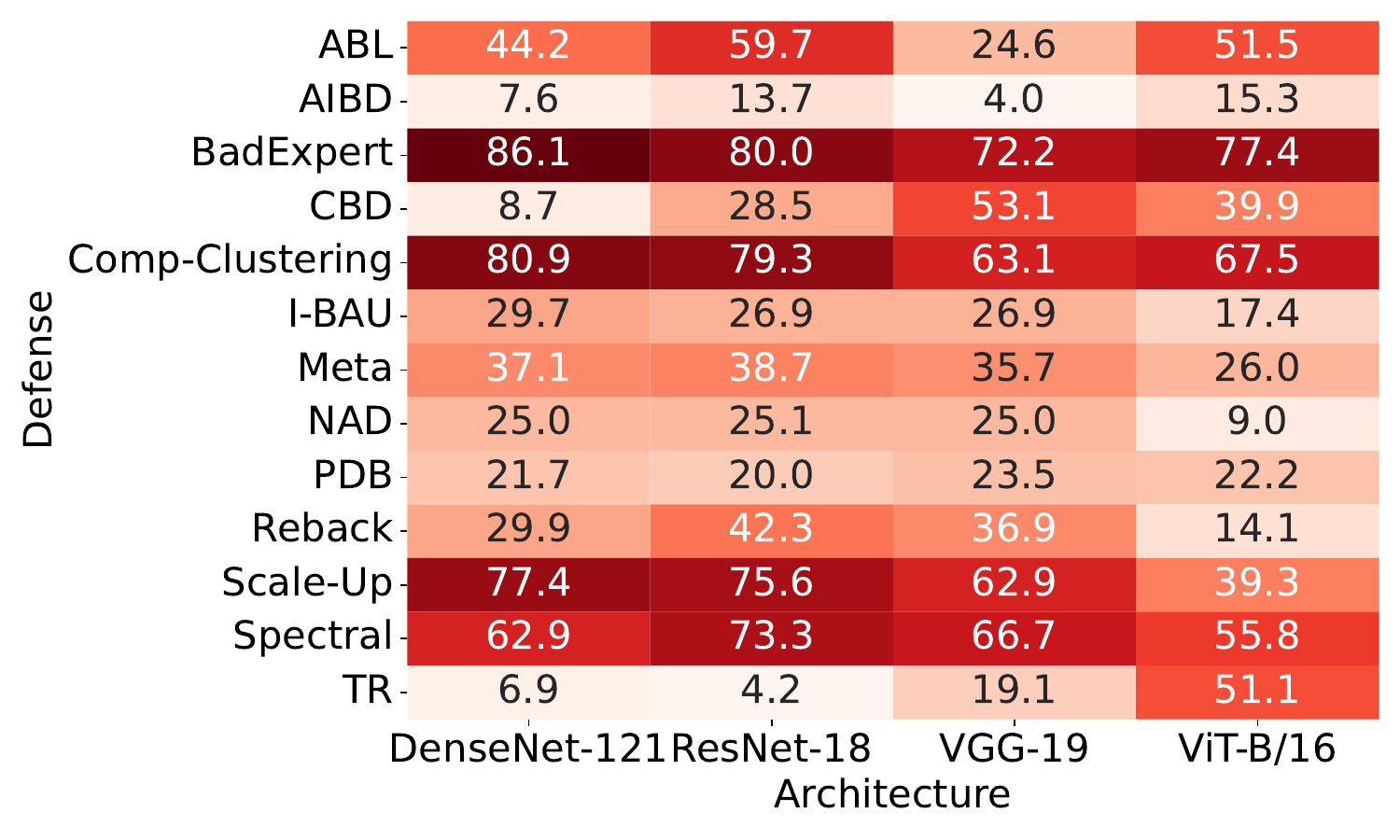}
        \caption{ASR (\%)}
    \end{subfigure}
    
    \begin{subfigure}[b]{0.8\columnwidth}
        \centering
        \includegraphics[width=\columnwidth, trim=10pt 16pt 10pt 10pt, clip]{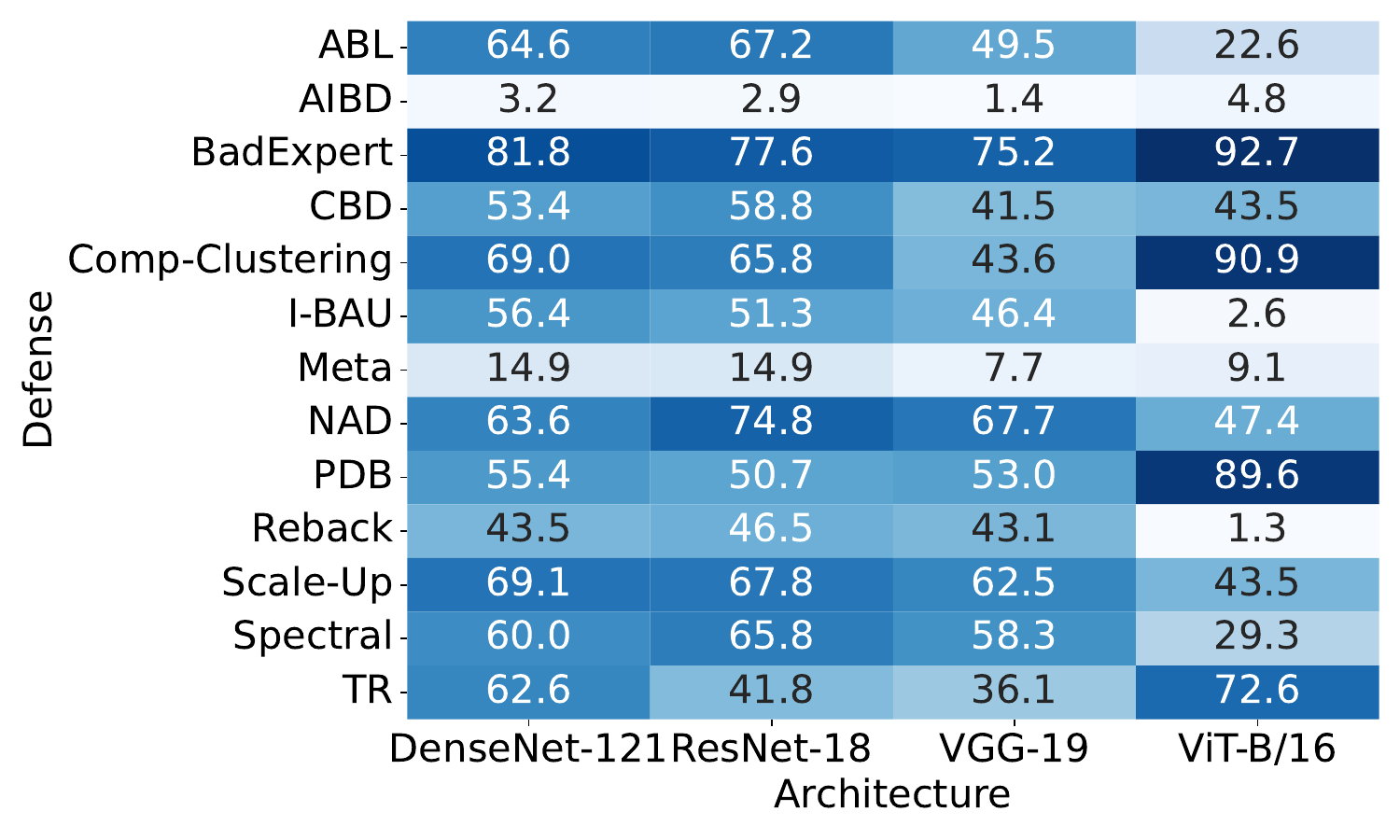}
        \caption{CA (\%)}
    \end{subfigure}
     \caption{Average CA and ASR for CIFAR-100 for different architectures.} 
    \label{fig:CIFARAvgArchitecture}
\end{figure}

\textbf{Observations.} Examining the results of defenses grouped and averaged by architecture across MNIST, CIFAR-100, and ImageNet-1K (Figures~\ref{fig:CIFARAvgArchitecture}, \ref{fig:MNISTAvgArchitecture}, and \ref{fig:ImagenetAvgArchitecture}) shows that the dominant pattern is that dataset complexity amplifies the CA–ASR trade-off: moving from MNIST to CIFAR-100 to ImageNet-1K, average CA drops sharply and many defenses fail on ImageNet-1K (several rows show CA $\approx 0$). On MNIST and CIFAR-100, PDB and NAD consistently deliver the best joint outcomes (high CA with low–moderate ASR), with PDB edging MNIST (e.g., CA $\approx99$, ASR $<10$) and NAD leading CIFAR-100 overall (CA in the mid-60s to mid-70s on CNNs with ASR $\approx 25$). In contrast, methods like BadExpert, Comp-Clustering, and Spectral often preserve CA, but the defense is not effective as it does not decrease the ASR. On ImageNet-1K, only Metasift scales, as it achieves near-zero ASR ($\leq 1$) while keeping high CA (roughly $66\text{–}80$) across all architectures. Everything else is either leaky (high ASR) or destructive (near-zero CA). 
We observe that for CIFAR-100, TR is excellent on DenseNet-121 (ASR $\approx 7$ with CA $\approx 63$), yet unstable elsewhere. Specifically, for ViT-B/16, the CA is kept relatively high, but the ASR drops only to $\approx 51$, illustrating CNN-specific tuning. Moreover, this is visible in the MNIST results (Figure~\ref{fig:MNISTAvgArchitecture}), as the CA is kept high for all CNN-based models, but is significantly negatively affected for many defenses with ViT-B/16. PDB and NAD are comparatively robust across CNNs (DenseNet/ResNet/VGG) on MNIST/CIFAR-100, while several methods (I-BAU, NAD, Reback) severely degrade CA on ImageNet regardless of model. Interestingly, ViT-B/16 is the weakest on CIFAR-100 for many defenses, but becomes the most usable on ImageNet when paired with Meta (high CA, ASR $\approx 0$), suggesting capacity/inductive-bias help only if the defense scales. Overall, defense effectiveness is tightly coupled with both architecture and dataset scale: CNNs benefit from spatially grounded, input- or parameter-level defenses, whereas ViTs require strategies that exploit global patterns to preserve CA while mitigating attacks.
\begin{challenge}
\label{chall: architecture agnostic}
 While aiming to be model- and dataset-agnostic, current defenses exhibit strong sensitivity to both. Future research should aim to design truly architecture-agnostic defense frameworks that generalize across model families and datasets.
\end{challenge}

\paragraph{\textbf{\textit{Trigger Types}}}

\begin{figure}
    
     \begin{subfigure}[b]{0.8\columnwidth}
        \centering
        \includegraphics[width=\columnwidth, trim=10pt 13pt 10pt 10pt, clip]{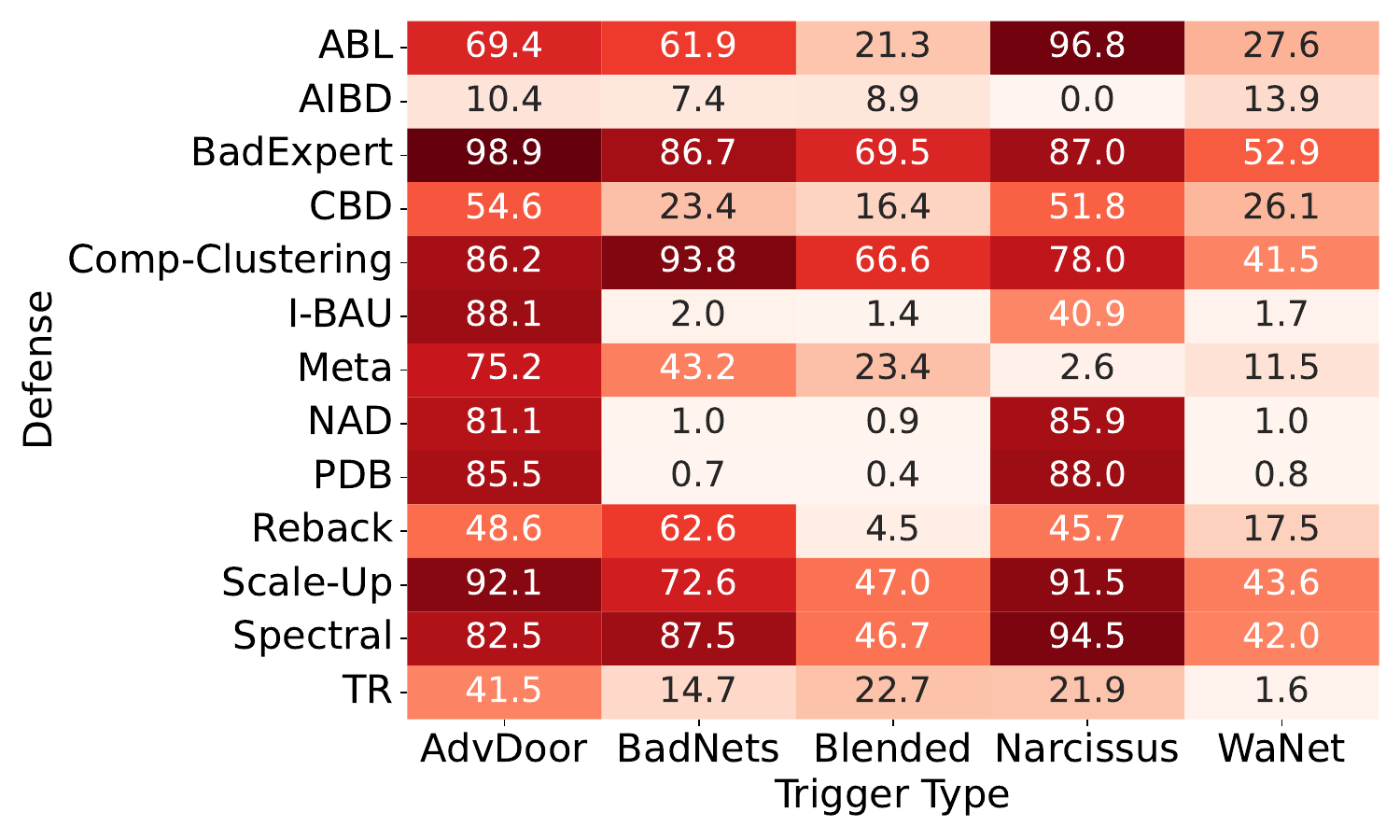}
        \caption{ASR (\%)}
        \label{fig:CIFARAVGTriggerTypeASR}
    \end{subfigure}
    \begin{subfigure}[b]{0.8\columnwidth}
        \centering
        \includegraphics[width=\columnwidth, trim=10pt 13pt 10pt 10pt, clip]{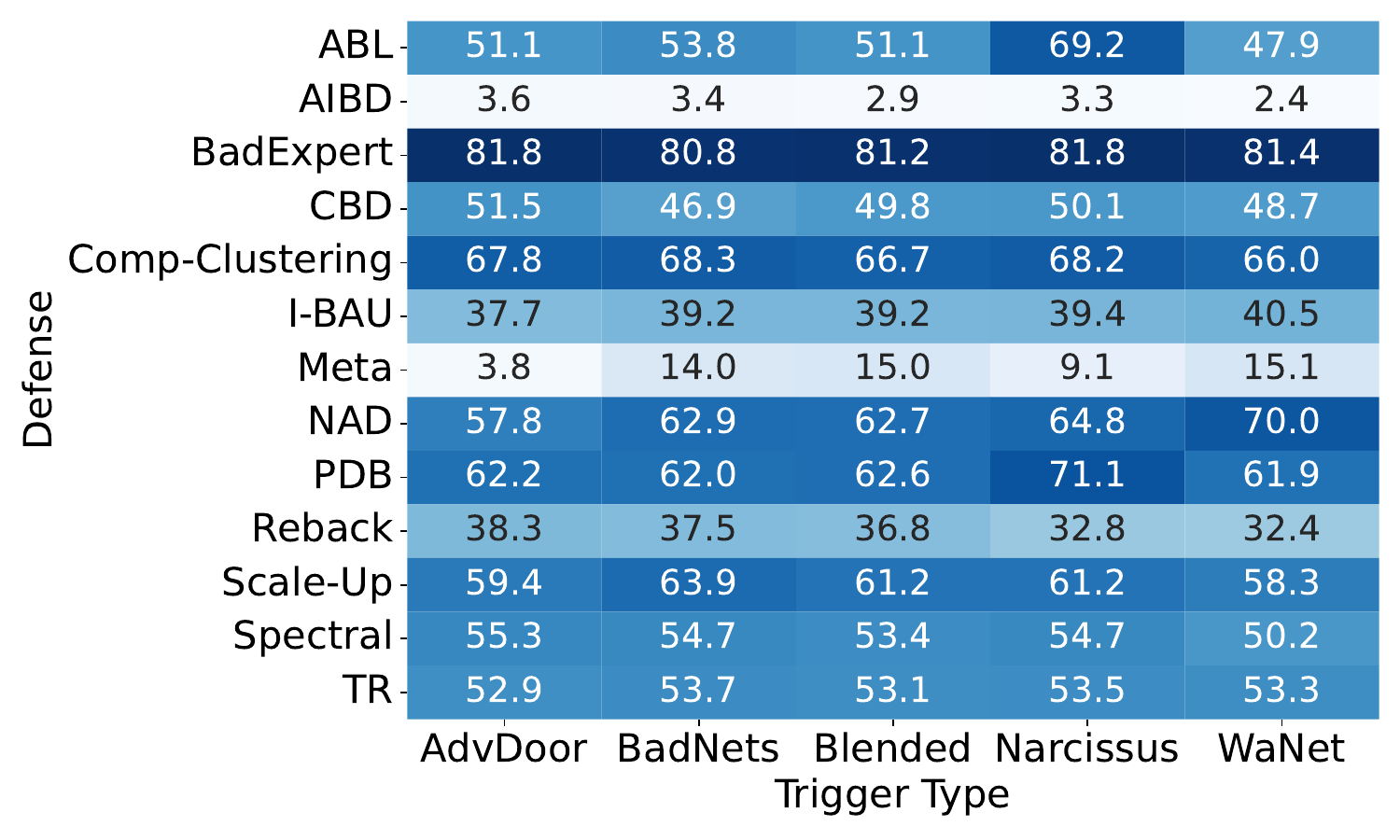}
        \caption{CA (\%)}
    \end{subfigure}
     \caption{Average CA and ASR for CIFAR-100 for different trigger types.} 
     \label{fig:CIFARAvgTriggerType}

\end{figure}

Based on the review in Section~\ref{sec:literature study}, most defenses are evaluated primarily against legacy, simple, and highly visible attacks. Only a few works consider advanced, stealthy triggers, such as adaptive or clean-label attacks, which are essential for realistic threat modeling. We also observe that some defenses underperform even in scenarios considered ``too simple'' to challenge them. The common assumption that a defense effective under worst-case conditions will also succeed in simpler ones does not always hold. For instance, in Figure~\ref{fig:CIFARAVGTriggerTypeASR}, the compatibility clustering fails against BadNets (ASR 93.8\%).

\begin{recommendation}
\label{rec: different difficulty level of attacks in evaluation}
    Defenses need to be evaluated under various difficulty conditions, including simpler attacks, but also worst-case scenarios and adaptive attacker settings.
\end{recommendation}

\textbf{Observations.} Analyzing defenses grouped and averaged by trigger types, Figures~\ref{fig:CIFARAvgTriggerType}, \ref{fig:MNISTAvgTriggerType}, and \ref{fig:ImagenetTriggerType}, reveal that on MNIST, the trigger choice (BadNets/Blended/WaNet) has little effect: NAD and PDB consistently achieve low ASR ($\approx10$–$18\%$) while keeping CA very high ($\approx92$–$99\%$), and I-BAU can push ASR lower on WaNet at the cost of CA. On CIFAR-100, trigger structure becomes decisive: localized/additive triggers remain tractable (PDB/NAD drive ASR to $\leq1\%$ with CA in the $60$–$70\%$ range), while semantically integrated, clean-label trigger (Narcissus) evade pixel/parameter-level inspection. Methods that minimize ASR on these hard triggers (e.g., AIBD, Metasift) do so by crushing CA to single digits, while TR yields a somewhat usable trade-off (e.g., Narcissus $\approx22\%$ ASR at $\approx53\%$ CA). Outlier-filtering approaches (Spectral, Scale-Up) often preserve CA but also leave high ASR, meaning that the backdoors remain active on the non-filtered samples.
On ImageNet-1K, defense scalability overwhelms trigger differences: Metasift uniquely sustains high CA ($\approx71\%$) with near-zero ASR ($\leq0.7\%$) across BadNets/Blended/WaNet, while alternatives either leak (e.g., BadExpert except on WaNet) or collapse CA (e.g., I-BAU, NAD, ABL). Note that missing entries indicate timeouts/failures, underscoring non-scalability. Overall, simpler, spatially bounded triggers are easier to detect, while semantically integrated or high-visibility triggers increasingly challenge defenses as model and dataset complexity grow. 

\begin{challenge}
\label{chall: generalize across trigger types}
    Existing defenses fail to generalize across spatial, semantic, or high-visibility patterns, especially on large-scale datasets.
\end{challenge}


\paragraph{\textbf{\textit{Poisoning Rates}}}



Section~\ref{sec:literature study} shows that evaluation practices lack standardization. Researchers often use arbitrary, typically low poisoning rates without a common benchmark, limiting cross-study comparability and biasing results toward defenses. Many also omit reporting attack success without defenses.

\begin{figure}
    \begin{subfigure}[b]{0.8\columnwidth}
        \centering
        \includegraphics[width=\columnwidth, trim=10pt 13pt 10pt 11pt, clip]{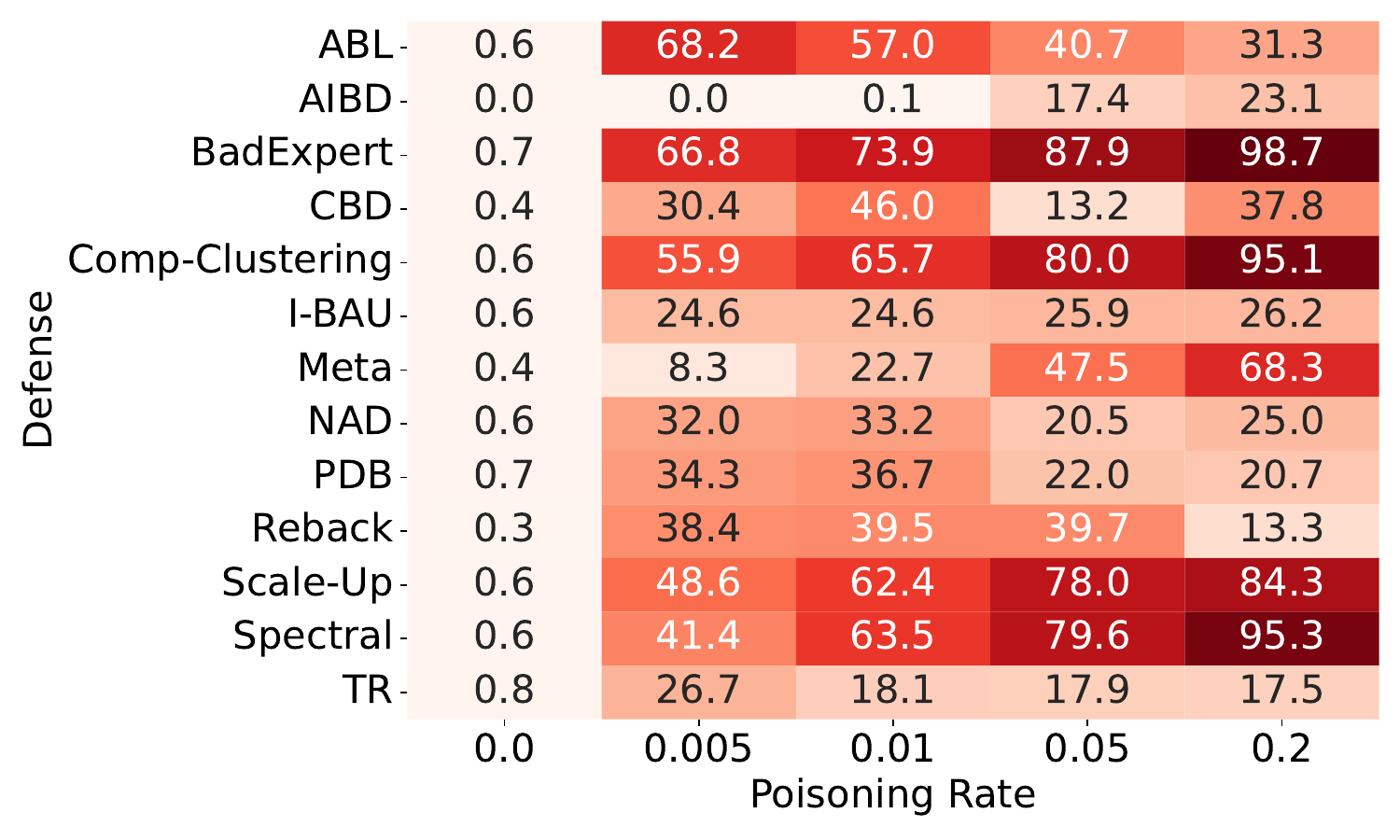}
        \caption{ASR (\%)}
    \end{subfigure}

    \begin{subfigure}[b]{0.8\columnwidth}
        \centering
        \includegraphics[width=\columnwidth, trim=10pt 13pt 10pt 10pt, clip]{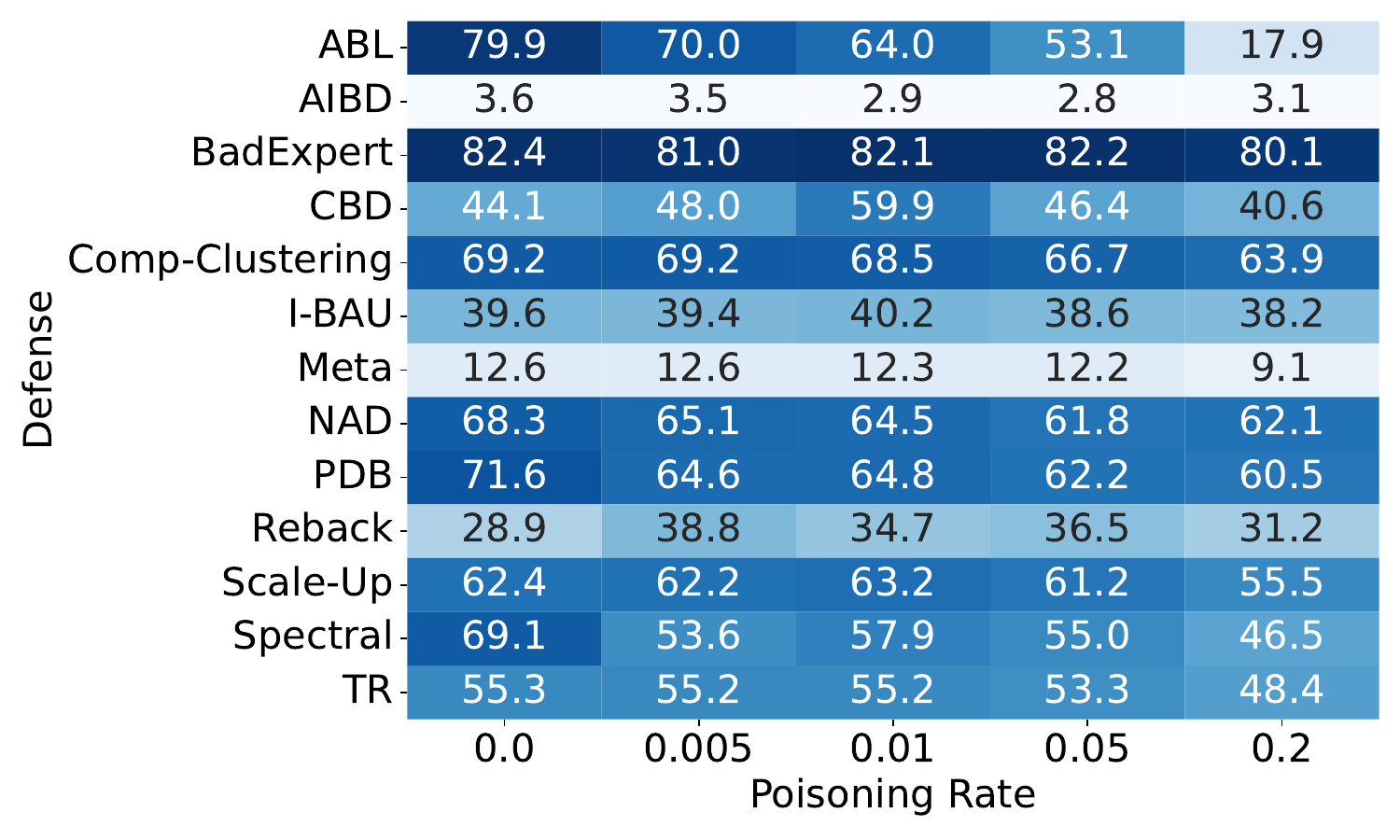}
        \caption{CA (\%)}
    \end{subfigure}
    \caption{Average CA and ASR for CIFAR-100 for different poisoning rates.}  
    \label{fig:CIFARAvgPoisningRate}
    
\end{figure}

\textbf{Observations.} In Figures~\ref{fig:CIFARAvgPoisningRate},~\ref{fig:MNISTAvgPoisningRate}, and~\ref{fig:ImagenetAvgPoisningRate} we examine defenses grouped and averaged by increasing poisoning rates, highlighting the sensitivity of methods to both dataset complexity and attack strength. On MNIST, low poisoning rates (0.005-0.05) are often mitigated by parameter-level defenses (e.g., PDB, CBD, and IBAU), preserving CA while reducing the ASR. Higher rates (0.2) degrade CA and allow elevated ASR, revealing limits of defenses when attacks become pervasive. On CIFAR-100, even modest poisoning significantly reduces defense effectiveness: many methods see sharp ASR increases and CA drops, especially for complex attacks and deeper models. Large-scale datasets like ImageNet-1K worsen this effect, where high-resolution and semantically diverse inputs make most defenses fragile against even low poisoning. It is interesting to see how, even in the absence of attacks, many defenses sensibly decrease the CA of the model, affecting its functionalities, giving further support for~\ref{rec: report defense performance in no-attack case}. The initial concept highlights that defenses, although designed to enhance robustness, can inadvertently harm normal performance. Overall, robustness declines nonlinearly with the poisoning rate, with measurable impacts appearing even without attacks, and the interaction between model and dataset scale determines how effectively a defense preserves performance.

\begin{challenge}
\label{chall: robust against diff poison rates}
    Ensuring consistent robustness of the defense across poisoning strengths without sacrificing clean performance remains an open challenge.
\end{challenge}

\paragraph{\textbf{\textit{Hyperparameter Selection}}}
Based on the review of the papers for this SoK and the re-implementation from the official repositories to perform experiments, the defenses are designed to work under different conditions, using different hyperparameters.
Most of them were estimated experimentally without providing a clear strategy to choose them in an unknown scenario. Thus, in our experiments, we used the default hyperparameters defined in their official code.

\begin{figure*}
        \centering
        \includegraphics[width=\textwidth, trim=10pt 18pt 10pt 11pt, clip]{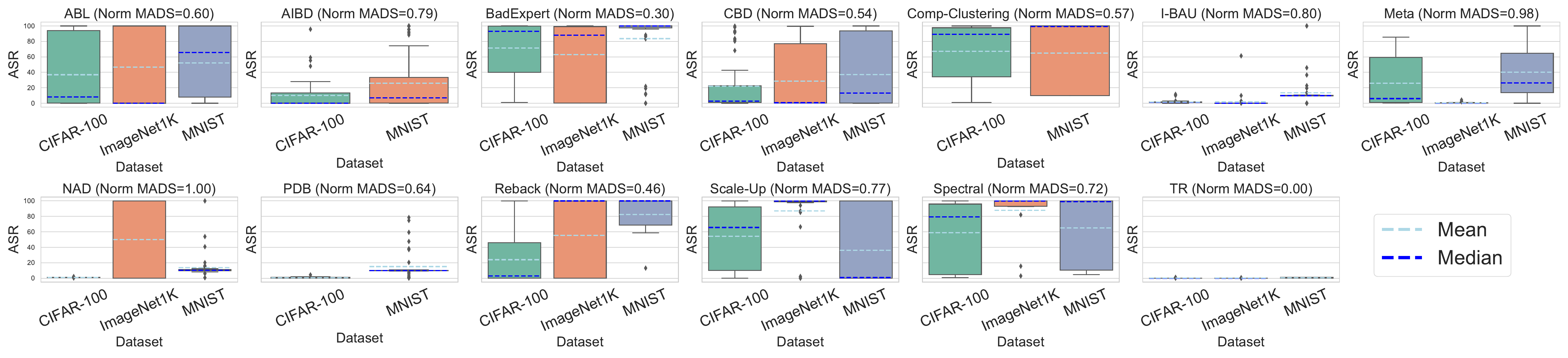}
        \caption{Comparison of the attack ASR on various datasets using default hyperparameters to different defenses.}
        \label{fig:hyperparametersDatasetComparisonASR}
\end{figure*}

\textbf{Observations.} To systematically quantify the adaptability of a defense across datasets, we introduce the Mean Absolute Difference Score (MADS). This metric measures how consistent the ASR of a defense remains when tested across multiple datasets under the same experimental settings. For a given defense $d$ and experiment $e$, let $\text{ASR}_{d,e}^{D_i}$ denote the ASR on dataset $D_i$, and let
\[
\mu_{d,e} = \frac{1}{|D|}\sum_{i=1}^{|D|} \text{ASR}_{d,e}^{D_i}
\]
be the mean ASR across all datasets $D$. The MADS for defense $d$ is then defined as the average mean absolute deviation across experiments:
\[
\text{MADS}_d = \frac{1}{EX\cdot|D|}\sum_{e=1}^{EX} \sum_{i=1}^{|D|} \big| \text{ASR}_{d,e}^{D_i} - \mu_{d,e} \big|,
\]
where $EX$ is the number of experiments.
A lower MADS indicates that a defense's ASR is more consistent across datasets, while a higher MADS indicates greater variability.
In Figure~\ref{fig:hyperparametersDatasetComparisonASR}, the results exhibit significant variability across datasets. This highlights that most defenses do not offer global protection when deployed in a scenario different from the one evaluated in their respective papers. The only defenses nearing zero are TR, I-BAU, and PDB. Examining the CA for the former two in the case of Imagenet-1K shows a complete reduction to (near) zero, while PDB completely fails with Imagenet-1K, highlighting a critical situation even in those that initially appeared more promising. 

\begin{challenge}
\label{chall:less sensitive hyperparameters}
    Design defenses that minimize hyperparameter sensitivity across datasets, improving defense generalizability.
\end{challenge}

\begin{recommendation}
\label{rec: tune on multiple datasets}
    Tune and validate defenses on multiple datasets to ensure consistent cross-dataset effectiveness.
\end{recommendation}

\paragraph{\textbf{\textit{Variability analysis}}}
\begin{figure}[!ht]
        \centering
        \includegraphics[width=\linewidth, trim=14pt 20pt 10pt 11pt, clip]{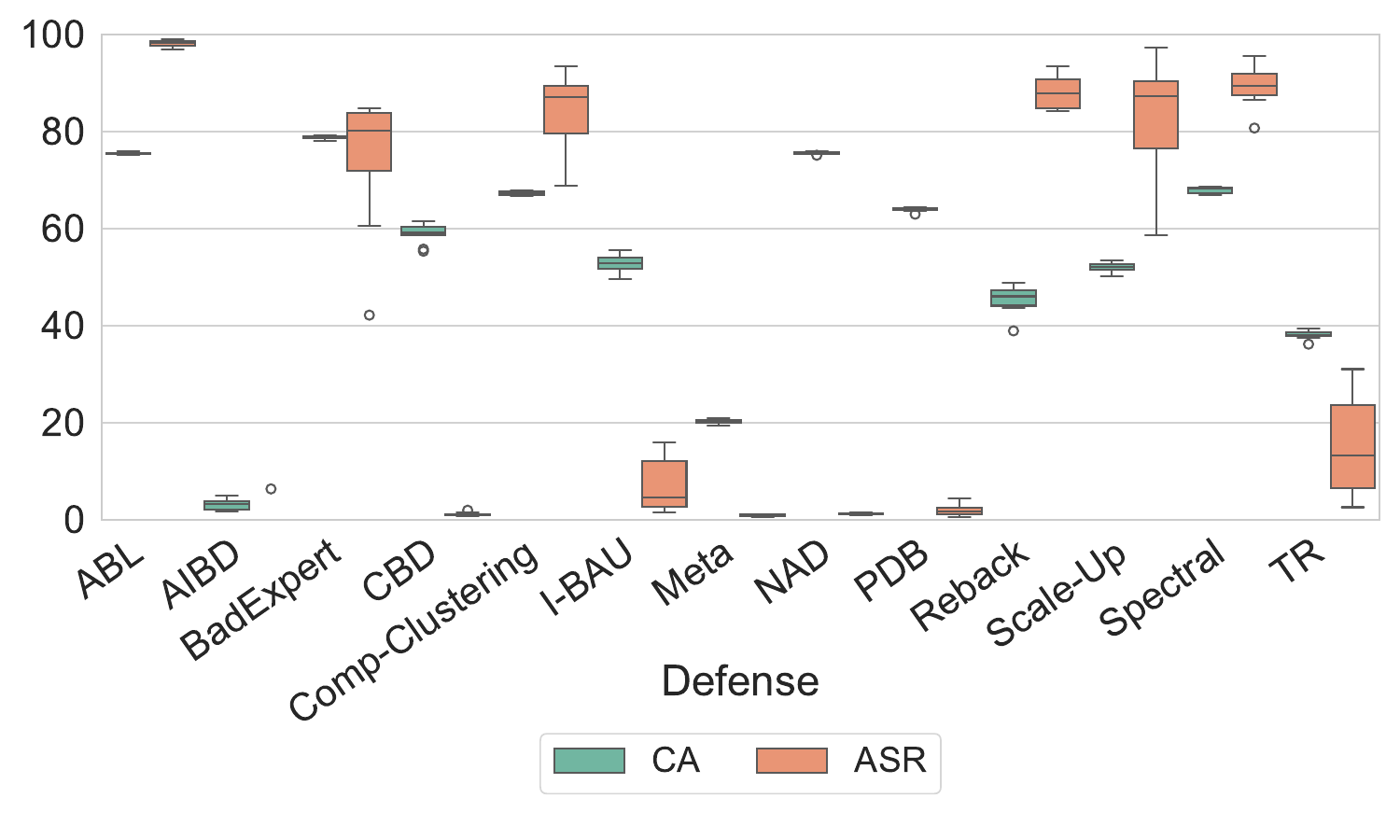}
        \caption{Performance consistency (CA and ASR) using BadNets against defense methods on CIFAR-100 with ResNet-18 (0.5\% poison rate). Each defense shows two distributions: CA (left, green) and ASR (right, red) computed over ten independent trials.}
        \label{fig:variability_plot}
\end{figure}

Repeated experiments on CIFAR-100 with ResNet-18 under the BadNets attack (0.5\% poison rate) reveal noticeable variability across defenses. The CA remains stable for most methods, while ASR shows high fluctuations on more than half of the evaluated defenses (see Figure~\ref{fig:variability_plot}). These results highlight the need to systematically measure variability and report average performance along with its dispersion (e.g., mean $\pm$ stdev, Coefficient of Variation, IQR). Reporting these results allows a precise assessment of defense stability across runs and ensures that results reflect consistent, reproducible behavior.

\begin{recommendation}
\label{rec:variability}
    Evaluations must include repeated runs to capture performance variability and report results using statistical measures such as the mean standard deviation ($\pm$), interquartile range (IQR), and coefficient of variation (CV).
\end{recommendation}

\paragraph{\textbf{\textit{Execution Time}}}

\begin{figure}[ht]
    \centering
    \begin{subfigure}[b]{0.49\linewidth}
        \centering
        \includegraphics[width=\linewidth]{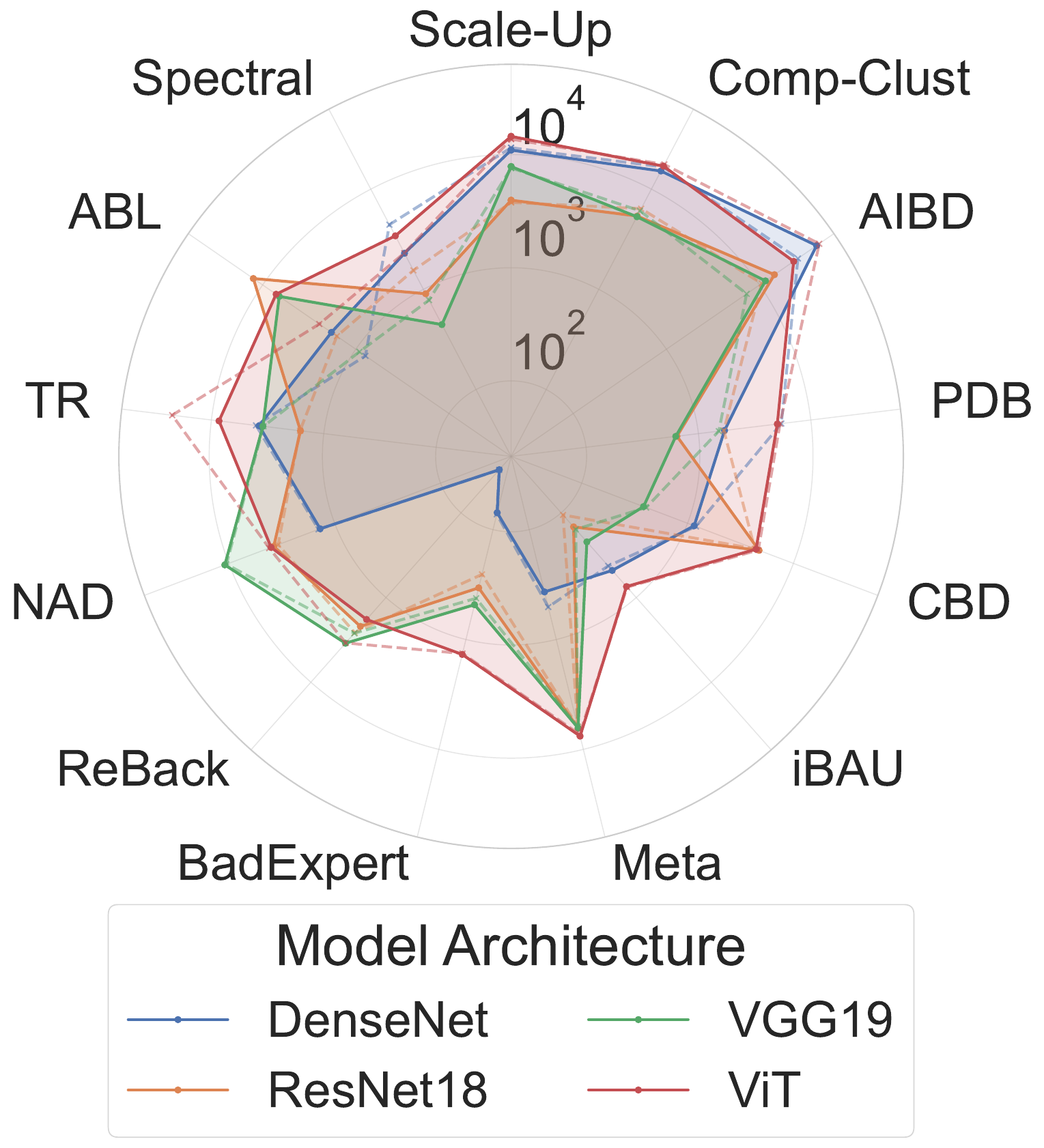}
        \caption{Grouped by architecture.}
    \end{subfigure}
    \hfill
    \begin{subfigure}[b]{0.49\linewidth}
        \centering
        \includegraphics[width=\linewidth]{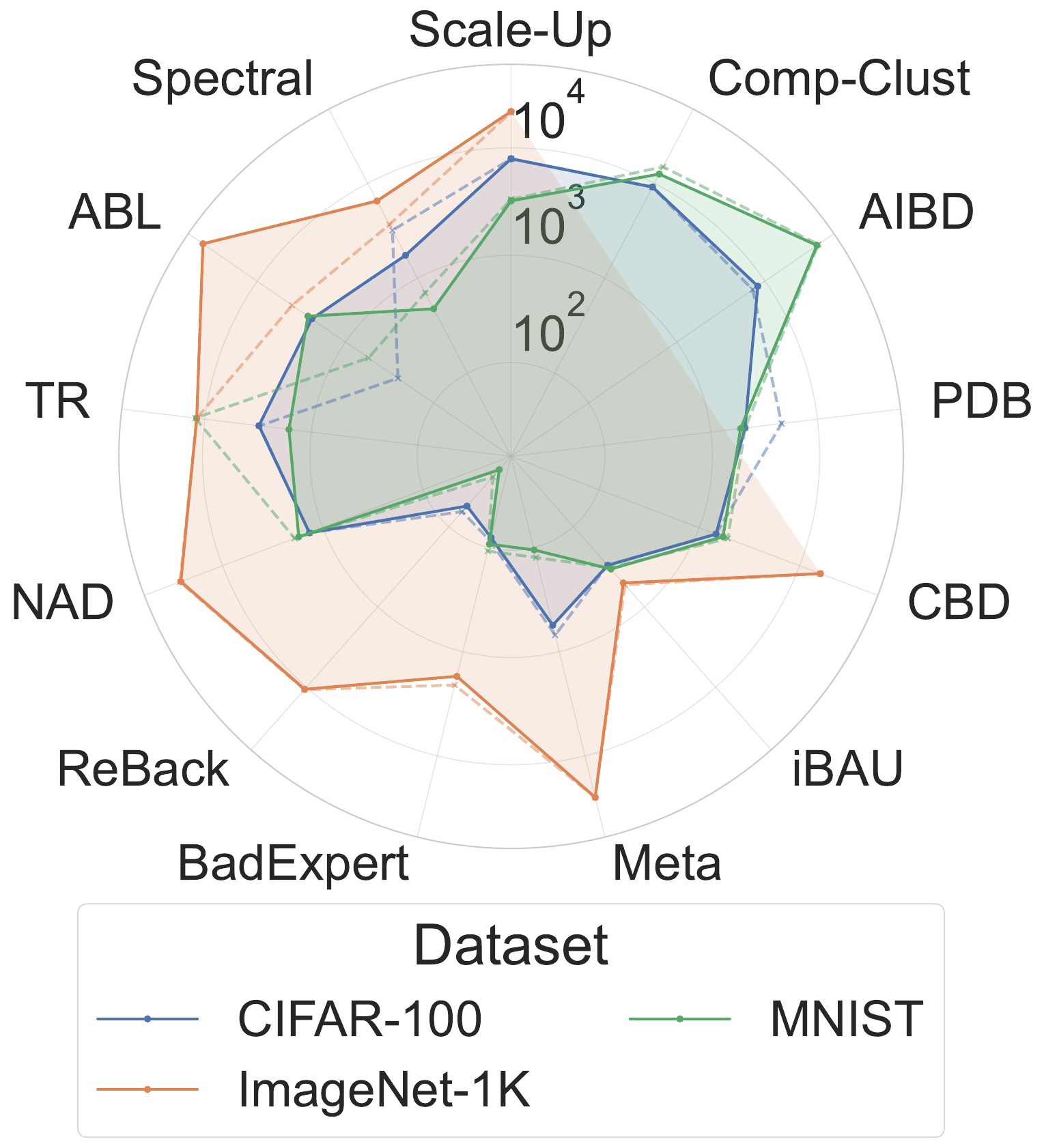}
        \caption{Grouped by dataset.}
    \end{subfigure}
    \caption{Average elapsed time (log scale) of defenses under different evaluation settings. Each radar plot summarizes the mean runtime across all attacks and poison rates. Dotted lines represent the runtime for each defense when no backdoor is present.}
    \label{fig:elapsed-time}
\end{figure}

The analysis in Section~\ref{sec:literature study} highlights a persistent lack of runtime reporting in defense evaluations. Execution time, while fundamental for assessing scalability and practicality, is often omitted or incompletely reported. Knowing how long a defense takes to operate, both in the presence and absence of a backdoor, is essential to evaluating its feasibility for real-world deployment.

\textbf{Observations.} In Figure~\ref{fig:elapsed-time}, we report the average elapsed time of all evaluated defenses across datasets and model architectures, comparing their runtime when a backdoor is present versus when it is not. The results show that runtime behavior is highly defense-dependent. Some methods, such as \textit{Spectral Signatures}, \textit{Reback}, and \textit{CBD}, exhibit minor variation between the two conditions, indicating stable and predictable execution cost. Others, however, display substantial differences: while certain defenses run faster on clean models, others become slower without a backdoor, likely due to convergence mechanisms or additional verification when no anomalies are detected. Moreover, several defenses, specifically \textit{Compatibility Clustering}, \textit{AIBD}, and \textit{PDB}, did not complete within a reasonable time frame (10 hours). These methods rely on extensive feature-space clustering or repeated gradient-based optimization procedures, which scale poorly with dataset size. These findings also further support~\ref{rec:report execution time}.

\begin{challenge}
    Achieving stable and efficient runtimes across defenses remains a critical challenge to scalability and real-world deployment.
\end{challenge}


\paragraph{\textbf{\textit{Case Study: Detection Defenses}}}

\begin{figure}[htpb]
        \centering
        \includegraphics[width=\linewidth,trim=7pt 8pt 10pt 8pt, clip]{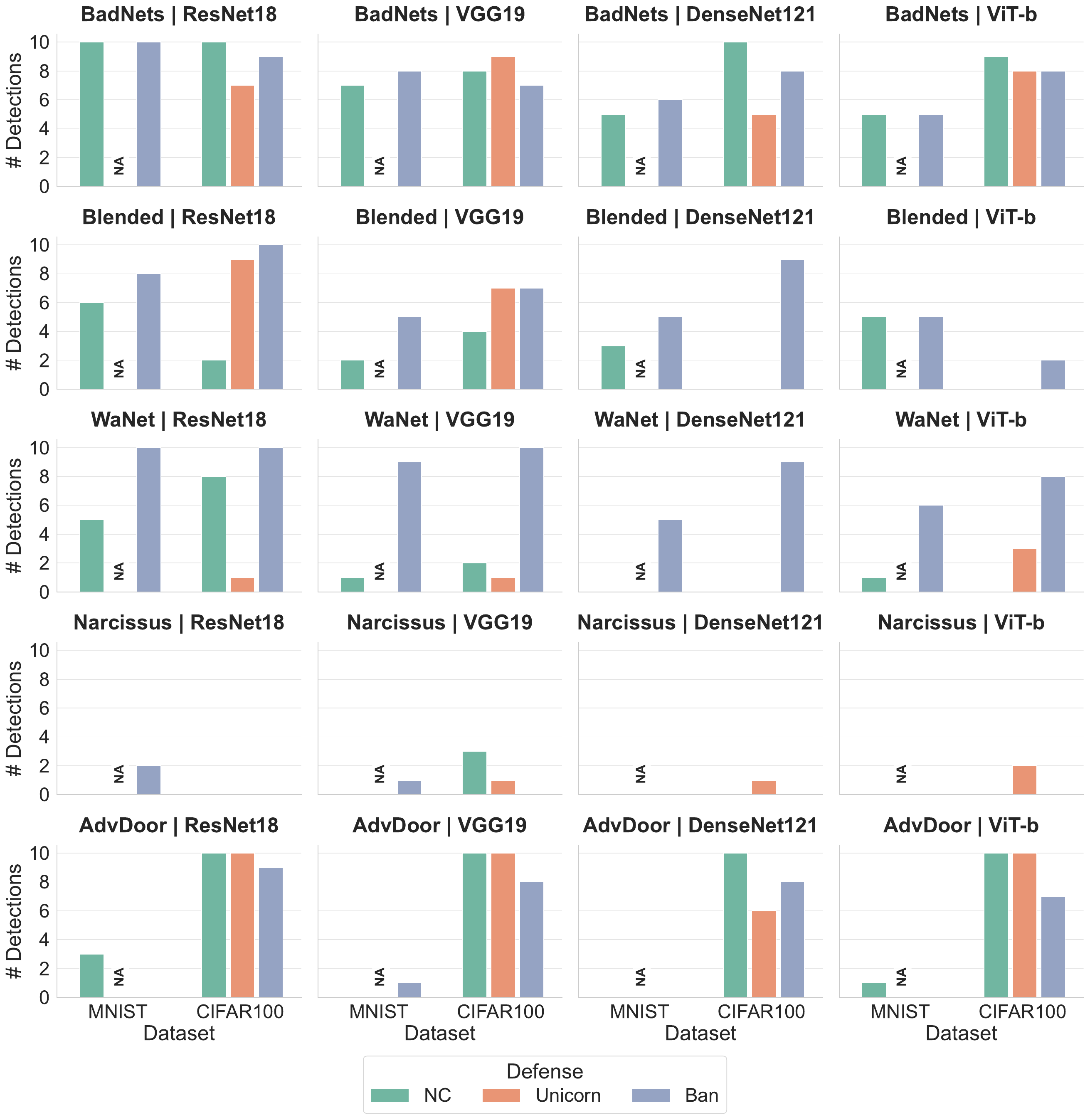}
        \caption{Performance of detection-based defenses on MNIST and CIFAR-100. We have no results for Unicorn and MNIST, as it could not be applied out of the box to one-dimensional images. 
        }
\label{fig:detection-def}
\end{figure}


We have also tested the chosen attacks against detection-based backdoor defenses like Neural Cleanse~\cite{wang2019neural}, Unicorn~\cite{DBLP:conf/iclr/WangMZM23}, and BAN~\cite{xu2024ban}. In this experiment, we poisoned 10 models and used the defenses to identify them. The results are shown in Figure~\ref{fig:detection-def}. Unicorn could not be applied to single-channel (grayscale) images without modification, so we did not run experiments with Unicorn on the MNIST dataset. The results vary depending on the chosen setup. For example, successful BadNets and AdvDoor attacks are detected by defenses across different models, whereas successful Narcissus evades them.\footnote{Note that for MNIST, Narcissus and AdvDoor attacks were not effective (low ASR), making low detection of backdoor reasonable.} On average, defenses work slightly better on ResNet18 than on other model architectures. Again, the trend of evaluating only against ResNet models, BadNets attack, and simpler datasets (CIFAR-10 and MNIST) does not fully support the generalizability of the evaluated defense. 

\subsection{Discussion on Sustainability}
\label{sec:sustainability}

As we reported in Section~\ref{sec:experiments}, to conduct, according to our estimates, a fair comparison among 15 defenses across more than 3\,000 experiments required more than 170 days of continuous computation. While this number, reflecting only the GPU time spent on the successful and reported experiments, already represents a considerable investment of resources, the actual resources consumed were substantially greater. The total experimental time in GPU hours was significantly extended by typical large-scale evaluation overheads, including repetitions due to experimental errors and necessary adjustments. Specifically, the reported experiments required about $78\%$ of the total available computing credits, while in reality, with the necessary implementation and preparation phase, the entire study consumed the full annual credit allocation for two accounts on the Netherlands national super-computing infrastructure (Snellius), primarily for the two large datasets, CIFAR-100 and ImageNet-1K. Several runs on Imagenet-1K frequently timed out after 10 (and 24) hours due to excessive resource consumption, while in some cases, memory consumption (over 300GB) led to failures. Moreover, experiments on MNIST were executed on an additional institutional compute cluster.
This highlights serious concerns regarding the sustainability and reproducibility of current defense evaluation practices. The growing number of backdoor defenses proposed in recent years, even within A* conferences, raises serious doubts about the realism and credibility of reported evaluations. This trend reflects broader pressure to publish rapidly, often at the expense of thorough, reproducible experimentation, leading to fragmented evaluation practices and limited comparability across studies.
At the same time, the lack of clear evaluation guidelines makes it difficult to design exhaustive, comparable setups, often leading researchers to rely on simple or easy-to-implement baselines and to run redundant experiments that waste valuable computational resources.
These observations reveal the need to establish standardized baselines that serve as reference leaderboards, enabling researchers to compare new defenses against verified results without repeating prior experiments, thereby promoting sustainable and environmentally responsible practices. We propose possible evaluation guidelines in Appendix~\ref{sec:guidelines}.

\begin{recommendation}
    Define standardized evaluation protocols to improve the efficiency and sustainability of defense evaluation.
\end{recommendation}



\section{Limitations}
\label{sec:limitations}

While we aimed to provide a detailed overview and experiments on backdoor attacks, we had to impose several constraints.
First, we limit our attention to the image domain. While the percentage of works focusing on the image domain justifies our choice, it would be interesting to consider other domains and to compare domain-agnostic defenses.
Next, we limit our attention to ``only'' nine venues. While those venues are commonly considered top-tier venues, there are more venues in the same category that we did not consider. Moreover, we do not consider journals. Finally, excellent defenses could be proposed in other venues (e.g., EuroS\&P, ACSAC, or SaTML), making our selection, while diverse, still far from exhaustive.

In our experiments, we consider 15 defenses, five attacks, four poisoning rates, four models, and three datasets. While this represents a considerable number of experiments, it can still be considered cherry-picked among all possible options. Indeed, since we list 183 defenses, evaluating only 15 means we skip many potentially excellent defenses.
Finally, to make our comparison as fair as possible, we do not tune the defenses' hyperparameters but use their default values. While we consider this option the only fair (and realistic) one, it may (and probably did) cause us to underestimate the potential performance of several defenses.

\section{Related Work}
\label{sec:related}


\begin{table}[t]
\caption{Scope comparison for related surveys and this SoK. Prior work provides high-level overviews of attacks and defenses, while our work systematically analyzes defense evaluation practices across multiple critical dimensions.}
\label{tab:comparison}
\centering
\resizebox{\columnwidth}{!}{%
\begin{tabular}{l*{10}{c}}
\toprule
\textbf{Paper} &
\rotatebox{90}{Metrics} &
\rotatebox{90}{Datasets} &
\rotatebox{90}{Models} &
\rotatebox{90}{Poisoning} &
\rotatebox{90}{Categorization} &
\rotatebox{90}{Trigger} &
\rotatebox{90}{Domains} &
\rotatebox{90}{Tasks} &
\rotatebox{90}{Exec. Time} &
\rotatebox{90}{Adaptive} \\
\midrule
Li et al.~\cite{li2022backdoor} & \checkmark & \checkmark & \checkmark & & \checkmark & & & & & \\
Gao et al.~\cite{gao2020backdoor} & & & & & & & & & & \\
Bai et al.~\cite{bai2024backdoor} & \checkmark & \checkmark & & & & & & & & \\
Shao et al.~\cite{li2023backdoor} & & & & & \checkmark & & & & & \\
BackdoorBench~\cite{wu2022backdoorbench} & \checkmark & \checkmark & \checkmark & & & & & & & \\
\midrule
\textbf{Our Work} & \checkmark & \checkmark & \checkmark & \checkmark & \checkmark & \checkmark & \checkmark & \checkmark & \checkmark & \checkmark \\
\bottomrule
\end{tabular}
}
\end{table}

Several surveys have systematically reviewed the literature, covering attack strategies, defensive mechanisms, and general taxonomies~\cite{li2022backdoor, gao2020backdoor, li2023backdoor, guo2022overview, bai2024backdoor}. While valuable, these works primarily focus on cataloging attack types and providing high-level defense categorizations rather than systematically analyzing how defenses are evaluated.
More structured evaluation efforts have emerged recently. BackdoorBench~\cite{wu2022backdoorbench} provides a unified benchmarking framework for comparing defenses under standardized settings. However, it evaluates only a small subset of defenses (12 methods) under limited threat models and does not address the broader methodological issues in defense evaluation across the literature. The rapid growth of backdoor defense research, with hundreds of papers published in recent years, makes ad-hoc comparisons increasingly difficult and highlights the need for systematic evaluation practices.

Table~\ref{tab:comparison} compares the scope of prior surveys against our work. Existing surveys cover high-level aspects such as attack categories, trigger types, and domains, but rarely detail experimental parameters like poisoning rates, execution-time constraints, or adaptive attacker considerations. Moreover, most do not explicitly categorize defenses in a way that enables reproducible cross-paper comparisons. In contrast, our work focuses exclusively on defenses, cataloging them along multiple dimensions (e.g., metrics, datasets, models, execution time, attacker adaptiveness) to facilitate fair and transparent benchmarking.
The works in Table~\ref{tab:comparison} highlight both attacks and defenses, explaining the general trends in backdoor attacks and defenses. These papers focus on a general explanation rather than comparing/analyzing the current defenses. The increasing number of publications makes comparing defenses (in a benchmark, reproducible, and fair manner) intractable. 

To our knowledge, this is the first SoK that focuses on defense evaluation in backdoor attacks. Rather than surveying attack or defense techniques, we systematically analyze how the community evaluates defenses across 183 papers from top-tier AI and security venues. We identify critical gaps in evaluation practices, including reliance on outdated attack methods, low-difficulty datasets, and weak threat models, among others, and provide actionable recommendations for rigorous, reproducible defense evaluation. Our analysis reveals that current evaluation practices may substantially overestimate defense effectiveness, with significant implications for deploying these techniques in real-world systems.

\section{Conclusions \& Future Work}
\label{sec:conclusions}

This work presents the first comprehensive systematization of backdoor defense evaluation. By analyzing 183 papers and conducting over 3\,000 experiments, this study offers a clear, data-driven perspective on dominant methodological trends and identifies key gaps within the existing literature. 
Our findings reveal that 73\% of defenses evaluate only on MNIST or CIFAR-10, only 58\% consider adaptive attackers, and execution time is rarely reported ($\approx5\%$). Empirically, we demonstrate that defense effectiveness varies dramatically across settings—methods that succeed on MNIST often fail on ImageNet-1K, and defenses robust to patch triggers struggle against more complex backdoors.
We provide concrete recommendations for evaluation: prioritize dataset diversity over quantity, test against both established and recent attacks (\ref{rec: diversity in evaluation}), report computational overhead (\ref{rec:report execution time}), and evaluate behavior under no attack conditions (\ref{rec: report defense performance in no-attack case}). Critical open challenges remain: developing architecture-agnostic defenses (\ref{chall: architecture agnostic}), achieving robustness across trigger types (\ref{chall: generalize across trigger types}), and designing hyperparameter-insensitive methods (\ref{chall:less sensitive hyperparameters}), to name a few. Our analysis---170 GPU-days for this study---underscores the need for standardized evaluation protocols to enable sustainable, reproducible research.

By releasing our framework and dataset, we aim to facilitate fair comparisons of defenses and guide practitioners in selecting appropriate defenses for real-world deployment. Future work should focus on establishing community-maintained leaderboards and developing theoretical frameworks that connect defense mechanisms to attack characteristics, rather than focusing on empirical results.

\appendix
\section*{Guidelines for Experimental Evaluation}
\label{sec:guidelines}

Based on our experimental analysis, we suggest a number of guidelines to follow:
\begin{compactitem}
    \item \textbf{Diversity over quantity}. Evaluate the performance of backdoor defense with different model families (e.g., CNN and ViT), dataset complexities (MNIST, CIFAR-100, and ImageNet-1K), and attack types (trigger in input-, feature-, and parameter-space). Compare with other state-of-the-art defenses with similar threat models or objectives.
    \item \textbf{Fair comparison}. Use the default hyperparameters for the new defense when comparing with defenses that use default hyperparameters. If the hyperparameters are tuned, provide the same tuning effort for each defense.
    \item \textbf{Result reporting}. Besides common metrics like ASR and CA, provide results on execution time and the model's behavior when defense is deployed, but no attack is present.
    \item \textbf{Statistical significance}. Provide multiple runs of experiments and report average and standard deviation values.
    \item \textbf{Adaptive attacker}. Always assess the performance of a defense in the presence of an adaptive attacker, which aligns with previous studies~\cite{carlini2019evaluating,arp2022and}, but is not always followed by the community.
    \item \textbf{Baselines}. Given the number of attacks and defenses proposed over the last few years, it is impossible to conduct a detailed experimental comparison (if for no other reason than computational constraints). As such, we see a need to maintain a public list of baseline results. Then, such arranged baselines must be reported in every paper. Of course, for this to be achieved, the community must agree that it is necessary, as it will require additional effort for experiments and to maintain the list. 
\end{compactitem}

\section*{Metrics}
\label{app:metrics}

\begin{figure}[!ht]
        \centering
        \includegraphics[width=\linewidth, trim=0pt 0pt 0pt 0pt, clip]{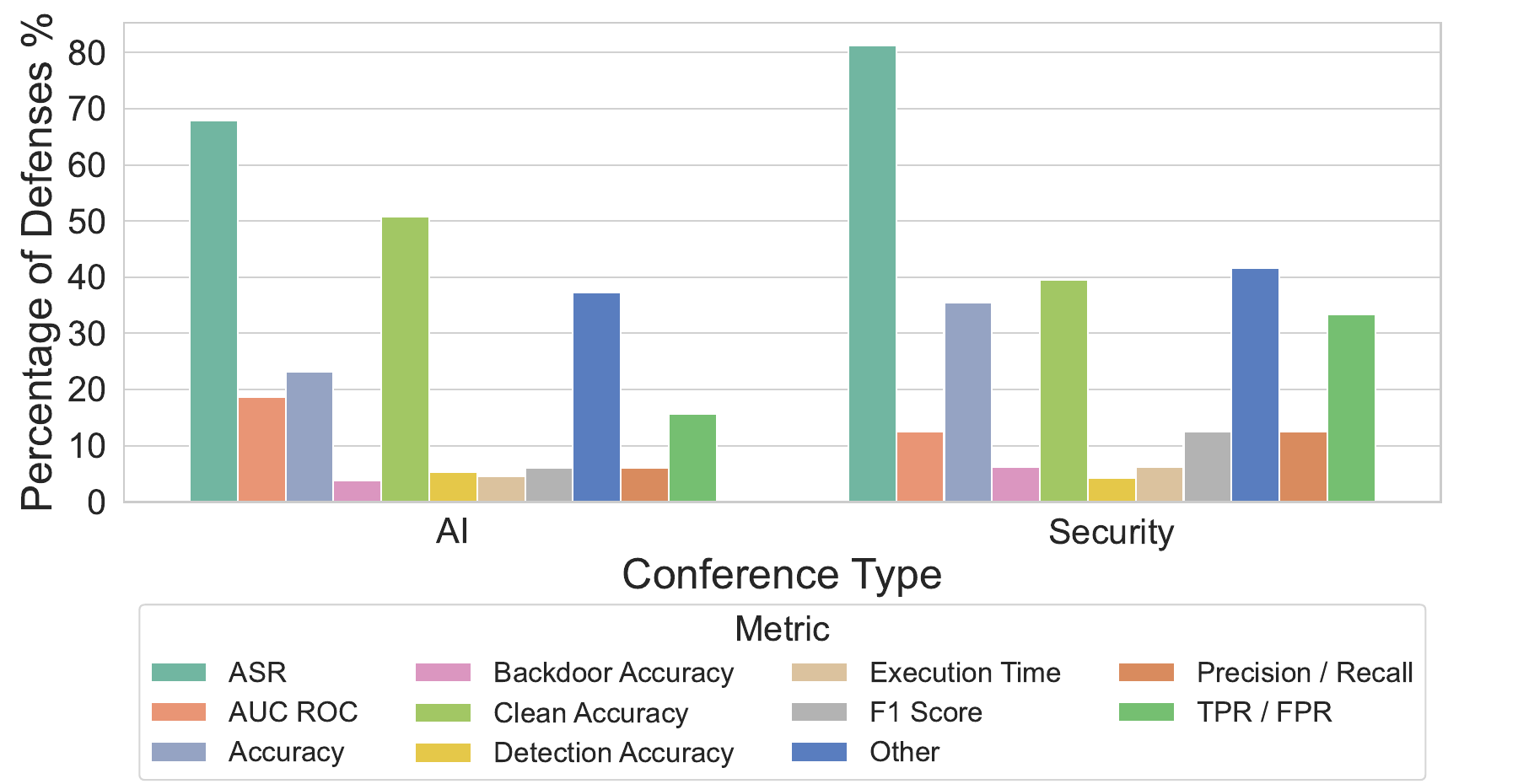}
        \caption{Usage patterns of evaluation metrics across defenses by conference type. ``Other'' includes metrics that were rarely reported and fall outside the ten most frequently used ones.}
        \label{fig:performance_metrics_by_conference_type_percentage}
\end{figure}

Figure~\ref{fig:performance_metrics_by_conference_type_percentage} shows the most used evaluation metrics for security and AI conferences. 

\section*{Analyzed Literature}

Table~\ref{tab:papers_evaluated} contains the full list of the evaluated papers.

\begin{table}
\scriptsize
\centering
\caption{List of analyzed papers per year and conference. ``/'' denotes no papers, while N/A means the conference has not happened yet.}
\label{tab:papers_evaluated}
\begin{tabular}{lll}
Year & Venue & Papers \\\toprule
\multirow{9}{*}{2025}    
& S\&P & \cite{DBLP:conf/sp/0001CLL0CH025, DBLP:conf/sp/ZhangZLLHLZ25, DBLP:conf/sp/Shen0000GY0AM025} \\
& NDSS & \cite{DBLP:conf/ndss/RiegerPKAKS25, DBLP:conf/ndss/ZhangBCM025, DBLP:conf/ndss/Yu0W00S025, DBLP:conf/ndss/ZengCP0DJ25, DBLP:conf/ndss/NguyenTJL25} \\
& ACM CCS & \cite{hailemariam2025poisonspot,yang2025filterfl,lachnit2025onhyperparameters}\\
& USENIX Security & \cite{DBLP:conf/uss/PopovicS0CK25} \\
& NeurIPS & N/A \\
& ICML & \cite{pmlr-v267-peng25f,pmlr-v267-min25b,pmlr-v267-he25v,pmlr-v267-niu25b,pmlr-v267-shen25d,pmlr-v267-ren25b,pmlr-v267-he25f}\\
& ICLR & \cite{DBLP:conf/iclr/YangTZCSW25, DBLP:conf/iclr/AlexSS025, DBLP:conf/iclr/Chen0HLCQ025, DBLP:conf/iclr/HuangELM025, DBLP:conf/iclr/WanSHZTY25, DBLP:conf/iclr/ZhangLXWDW25, DBLP:conf/iclr/YiHCLLCL25, DBLP:conf/iclr/YuanZWLW25, DBLP:conf/iclr/HeYZ0HY025} \\ 
& AAAI & \cite{DBLP:conf/aaai/ZhaoW25, DBLP:conf/aaai/0001H0V25, DBLP:conf/aaai/YinWLLL25, DBLP:conf/aaai/LiCZHCLS25, DBLP:conf/aaai/LinLLX25} \\
& CVPR & \cite{li2025psbd, DBLP:conf/cvpr/XuZH25, DBLP:conf/cvpr/HouLY25, DBLP:conf/cvpr/ZhangWYZZHHLZZ25} \\\midrule
\multirow{9}{*}{2024}    
& S\&P & \cite{DBLP:conf/sp/WangXMK24, DBLP:conf/sp/KabirSRM24, DBLP:conf/sp/MaYLYLLQ24, DBLP:conf/sp/MoZZLSHGX24, DBLP:conf/sp/ChengSTZZAXLMZ24} \\
& NDSS & \cite{DBLP:conf/ndss/Pei000S24, DBLP:conf/ndss/RiegerKMDS24, DBLP:conf/ndss/WeiM00ZFWZC24, DBLP:conf/ndss/FereidooniPRDS24} \\
& ACM CCS & \cite{yang2024distributed} \\
& USENIX Security & \cite{DBLP:conf/uss/LiD24, DBLP:conf/uss/0005RG24, DBLP:conf/uss/Sun0KS24, DBLP:conf/uss/KraussSD24} \\
& NeurIPS & \cite{DBLP:conf/nips/LinLWLX24, DBLP:conf/nips/MaiYP24, DBLP:conf/nips/WeiZW24, DBLP:conf/nips/XuFLTT24, DBLP:conf/nips/XuLKYP24, DBLP:conf/nips/MinQZ0C24, DBLP:conf/nips/PanYLSZK024, DBLP:conf/nips/HuangYSWL024, DBLP:conf/nips/YangJYHLXW024, DBLP:conf/nips/PooladzandiBJBP24, DBLP:conf/nips/XuG0JL24, DBLP:conf/nips/Mirzaei0NNMR0MS24} \\
& ICML & \cite{DBLP:conf/icml/XieFG24, DBLP:conf/icml/LiuXHY24, DBLP:conf/icml/MoHLL024, DBLP:conf/icml/YangGM24, DBLP:conf/icml/ZhaoXY24, DBLP:conf/icml/GaoC0LLS24, DBLP:conf/icml/HouFH0ZL24, DBLP:conf/icml/LiCC00W024, DBLP:conf/icml/Yuan000S24} \\
& ICLR & \cite{DBLP:conf/iclr/0005B24, DBLP:conf/iclr/XuH0Q024, DBLP:conf/iclr/PalY0S024, DBLP:conf/iclr/XieQHLWM24} \\ 
& AAAI & \cite{DBLP:conf/aaai/ZhaoL024, DBLP:conf/aaai/ChenW024a, DBLP:conf/aaai/QinCZYD24, DBLP:conf/aaai/ZhuNLXW24, DBLP:conf/aaai/Wang0XXAW24, DBLP:conf/aaai/ZhouLLMCM24, DBLP:conf/aaai/AnC0X0S0MCHZ24}\\
& CVPR & \cite{guan2024backdoor, li2024nearest} \\\midrule
\multirow{9}{*}{2023}    
& S\&P & \cite{DBLP:conf/sp/CaoJZG23, DBLP:conf/sp/ZhuTTWT23, DBLP:conf/sp/WeberXKZL23, DBLP:conf/sp/GongCYWGHS23, DBLP:conf/sp/KumariRFJS23} \\
& NDSS & \cite{DBLP:conf/ndss/ChuGISL23, DBLP:conf/ndss/MaWSXWX23, DBLP:conf/ndss/0005TLAX0S0XM023} \\
& ACM CCS & \cite{xie2023unraveling} \\
& USENIX Security & \cite{DBLP:conf/uss/PanZLLJ23, DBLP:conf/uss/QiXWWMM23, DBLP:conf/uss/Fu0J0LFY23, DBLP:conf/uss/ZengPJ0LJ23, DBLP:conf/uss/WangZW000LW023} \\
& NeurIPS & \cite{huang2023lockdown, DBLP:conf/nips/XiangXL23, DBLP:conf/nips/XiDLPJCMW23, DBLP:conf/nips/Yan000CSZ23, DBLP:conf/nips/0002F23, DBLP:conf/nips/MinQ0C23, DBLP:conf/nips/WeiZZW23, DBLP:conf/nips/YangGM23, DBLP:conf/nips/ZhuWZW23, DBLP:conf/nips/Chen000S23, DBLP:conf/nips/ShiDWGSL23, DBLP:conf/nips/0001CZMZ023, DBLP:conf/nips/TangYLLCH23, DBLP:conf/nips/HuangHCIT023, DBLP:conf/nips/JiaYSNRRLP23, DBLP:conf/nips/XianWSKBH023, DBLP:conf/nips/Shen000LAM023} \\
& ICML & \cite{DBLP:conf/icml/ZhuRC23, DBLP:conf/icml/XiangXL23, DBLP:conf/icml/LiLMKLLJ23, DBLP:conf/icml/KhaddajLMGSIM23}\\
& ICLR & \cite{DBLP:conf/iclr/JinSR23, DBLP:conf/iclr/WangMZM23, DBLP:conf/iclr/GuoLCG0023, DBLP:conf/iclr/0002TX0AL0SCM023} \\ 
& AAAI & \cite{DBLP:conf/aaai/YanWYL23, DBLP:conf/aaai/DoanLY023, DBLP:conf/aaai/LiuCMER23, DBLP:conf/aaai/00010WYYZ23, DBLP:conf/aaai/SunLM0L0Z23}\\
& CVPR & \cite{feng2023detecting, sun2023single, pang2023backdoor, mu2023progressive, liu2023detecting, tejankar2023defending, zhang2023backdoor} \\\midrule
\multirow{9}{*}{2022}    
& S\&P & / \\
& NDSS & \cite{DBLP:conf/ndss/RiegerNMS22} \\
& ACM CCS & / \\
& USENIX Security & \cite{DBLP:conf/uss/ShanB0Z22, DBLP:conf/uss/NguyenRCYMFMMMZ22} \\
& NeurIPS & \cite{bharti2022provable, DBLP:conf/nips/ChaiC22, DBLP:conf/nips/ChenWW22, DBLP:conf/nips/WangDZM22, DBLP:conf/nips/ZhuQCCZFD0WWS022, DBLP:conf/nips/LiuYM22, DBLP:conf/nips/CaiZCCW22, DBLP:conf/nips/WangHZZW22, DBLP:conf/nips/WangMDZM22, DBLP:conf/nips/ZhengTL022} \\
& ICML & \cite{DBLP:conf/icml/ShenLTX0AM022}\\
& ICLR & \cite{DBLP:conf/iclr/XiangMK22, DBLP:conf/iclr/GuoLL22, DBLP:conf/iclr/0002LCYJ022, DBLP:conf/iclr/Huang0WQ022, DBLP:conf/iclr/ZengCPM0J22} \\ 
& AAAI & \cite{DBLP:conf/aaai/JiaLCG22}\\
& CVPR & \cite{tao2022better, liu2022complex, gao2023backdoor, guan2022few} \\\midrule
\multirow{9}{*}{2021}    
& S\&P & / \\
& NDSS & \cite{shejwalkar2021manipulating, cao2020fltrust} \\
& ACM CCS & / \\
& USENIX Security & \cite{DBLP:conf/uss/Tang0TZ21, DBLP:conf/uss/AziziTWMPJRV21} \\
& NeurIPS & \cite{zheng2021topological, DBLP:conf/nips/WuW21, DBLP:conf/nips/LiLKLLM21, DBLP:conf/nips/SunLDHCL21} \\
& ICML & \cite{DBLP:conf/icml/abs21, DBLP:conf/icml/Xie0CL21, DBLP:conf/icml/ShenLTAX0M021}\\
& ICLR & \cite{DBLP:conf/iclr/LiLKLLM21, DBLP:conf/iclr/0001F21} \\ 
& AAAI & \cite{DBLP:conf/aaai/OzdayiKG21, DBLP:conf/aaai/ShanthamalluTS21}\\
& CVPR & \\\midrule
\multirow{9}{*}{2020}    
& S\&P &  / \\
& NDSS & / \\
& ACM CCS & / \\
& USENIX Security & /\\
& NeurIPS & / \\
& ICML & \cite{DBLP:conf/icml/RosenfeldWRK20}\\
& ICLR & \cite{DBLP:conf/iclr/DuJS20} \\ 
& AAAI & / \\
& CVPR & \cite{kolouri2020universal} \\\midrule
\multirow{9}{*}{2019}    
& S\&P & \cite{DBLP:conf/sp/WangYSLVZZ19} \\
& NDSS & / \\
& ACM CCS & \cite{liu2019abs} \\
& USENIX Security & /\\
& NeurIPS & \cite{DBLP:conf/nips/QiaoYL19} \\
& ICML & / \\
& ICLR & / \\ 
& AAAI & / \\
& CVPR & / \\\midrule
\multirow{9}{*}{2018}    & S\&P & / \\
& NDSS & / \\
& ACM CCS & / \\
& USENIX Security & / \\
& NeurIPS & \cite{DBLP:conf/nips/Tran0M18} \\
& ICML & / \\
& ICLR & / \\ 
& AAAI & / \\
& CVPR & / \\\bottomrule
\end{tabular}
\end{table}

\section*{Attacks Details}

In Table~\ref{tab:attack_summary}, we provide additional details on the attacks selected for our evaluation setting.
\begin{table}[t]
\caption{Summary of utilized backdoor attacks.}
\label{tab:attack_summary}
\centering
\small
\resizebox{\columnwidth}{!}{%
\begin{tabular}{@{}p{2.8cm}p{2.5cm}p{1.2cm}p{1.5cm}p{3.5cm}@{}}
\toprule
\textbf{Attack} & \textbf{Venue} & \textbf{Year} & \textbf{Type} & \textbf{Trigger Type} \\
\midrule
BadNets~\cite{gu2019badnets} & IEEE Access & 2017 & Dirty & Patch-based (static) \\
Blended~\cite{chen2017targeted} & arXiv & 2017 & Dirty & Image blending (transparent overlay) \\
WaNet~\cite{nguyen2021wanet} & IEEE S\&P & 2021 & Dirty & Spatial warping (imperceptible) \\
AdvDoor~\cite{zhang2021advdoor} & AAAI & 2022 & Dirty & Universal adversarial perturbation \\
Narcissus~\cite{zeng2023narcissus} & CVPR & 2023 & Clean & Sample-specific \\
\bottomrule
\end{tabular}
}
\end{table}
In Table~\ref{tab:avg_results_raw}, we report the mean CA and ASR for the considered backdoor attacks and for clean training without an attack. These provide baselines for fair comparison and analysis in Section~\ref{sec:experiments}. 
\begin{table*}[tb]
\centering
\caption{Average CA/ASR(\%) without defense.}
\label{tab:avg_results_raw}
\resizebox{0.85\textwidth}{!}{%
\begin{tabular}{llcccccc}
\toprule
   Dataset &   Model &       AdvDoor &       BadNets &       Blended &     Narcissus &         WaNet &   No Attack \\
\midrule
\multirow{4}{*}{MNIST} & DenseNet-121 &  99.21 / 4.65 & 99.18 / 99.99 & 99.25 / 100.0 &  99.34 / 0.08 & 76.78 / 99.92 & 99.14 / N/A \\
  &    ResNet-18 &  98.94 / 5.85 & 99.07 / 99.91 & 99.03 / 100.0 &  99.05 / 0.07 & 98.84 / 56.41 &  98.9 / N/A \\
  &       VGG-19 &  99.01 / 2.74 & 99.15 / 99.98 & 99.09 / 99.99 &   99.1 / 0.13 & 97.76 / 99.33 & 98.96 / N/A \\
  &     ViT-B/16 &  97.96 / 2.84 & 98.02 / 99.96 & 98.08 / 99.96 &   98.1 / 0.17 & 97.72 / 54.36 & 98.06 / N/A \\
\midrule
 \multirow{4}{*}{CIFAR-100} & DenseNet-121 & 78.51 / 99.96 & 70.39 / 99.33 & 69.43 / 49.62 & 78.97 / 82.39 & 68.96 / 45.29 & 79.09 / N/A \\
  &    ResNet-18 &  76.1 / 99.22 & 68.52 / 97.41 & 66.41 / 51.24 & 76.56 / 83.96 & 67.05 / 47.87 & 76.87 / N/A \\
  &       VGG-19 & 71.44 / 99.73 &  71.8 / 98.04 & 54.42 / 37.02 & 72.05 / 61.29 & 51.65 / 34.15 & 72.44 / N/A \\
  &     ViT-B/16 & 91.12 / 95.76 &  91.2 / 98.93 & 91.19 / 99.46 & 91.05 / 47.19 &  90.56 / 45.5 & 91.09 / N/A \\
 \midrule
\multirow{4}{*}{ImageNet1K} & DenseNet-121 &  69.65 / 0.02 & 70.48 / 79.82 & 69.41 / 89.79 &  69.52 / 0.99 & 69.07 / 68.14 & 69.57 / N/A \\
 &    ResNet-18 &  64.21 / 0.02 & 65.15 / 79.66 & 64.08 / 84.67 &  64.25 / 0.72 & 63.77 / 49.01 & 57.72 / N/A \\
 &       VGG-19 &  67.87 / 0.07 & 68.44 / 83.28 & 67.48 / 87.16 &  67.54 / 1.72 & 67.62 / 49.11 &  67.7 / N/A \\
 &     ViT-B/16 &   82.34 / 0.0 & 81.01 / 79.85 & 82.27 / 90.78 &  82.44 / 0.22 & 82.12 / 47.91 & 82.41 / N/A \\
     
\bottomrule
\end{tabular}}
\end{table*}

\section*{Model Type Groups}
\label{sec:model type groups}
In Table~\ref{tab:model_groups}, we provide examples of the grouping we used to separate different model types for analysis.

\begin{table}[tb]
\caption{Classification of neural network models into architecture families for systematic analysis.}
\label{tab:model_groups}
\resizebox{\columnwidth}{!}{%
\begin{tabular}{@{}ll@{}}
\toprule
NN Group Type & Example Variants to Include \\ \midrule
ResNet & ResNet, ResNet-18, ResNet-34, ResNet-50, ResNet-101, WideResNet, PreActResNet \\
VGG & VGG, VGG-11, VGG-16, VGG-19 \\
ViT & ViT, DeiT, CLIP-ViT, Swin Transformer \\
CNN & CNN, ConvNet, Custom CNN \\
GCN & GCN, GAT, GraphSAGE, GIN, RGCN, GatedGCN \\
MobileNet & MobileNet, MobileNetV2 \\
Transformer & BERT, RoBERTa, GPT, LLaMA, ALBERT, DistilBERT, GPT-2, GPT-3, GPT-4, T5 \\
DenseNet & DenseNet, DenseNet-121, DenseNet-169, DenseNet-161 \\
LeNet & Inception, GoogLeNet, LeNet, LeNet-5 \\
AlexNet & AlexNet \\
LSTM/GRU & LSTM, GRU, Bi-LSTM \\
MLP & MLP, FC-NN \\
Diffusion & DDPM, EDM, Stable Diffusion, DDIM \\
Reinforcement Learning & PPO, DRL, Policy Learning NN \\
Detection Models & Faster R-CNN, YOLO, SSD, DETR \\ \bottomrule
\end{tabular}
}
\end{table}

\section*{Additional Results}
\label{sec:additional_results}

\begin{figure}[tbp]
    \centering
    \begin{subfigure}[b]{0.8\linewidth}
        \centering
        \includegraphics[width=\linewidth, trim=10pt 13pt 10pt 10pt, clip]{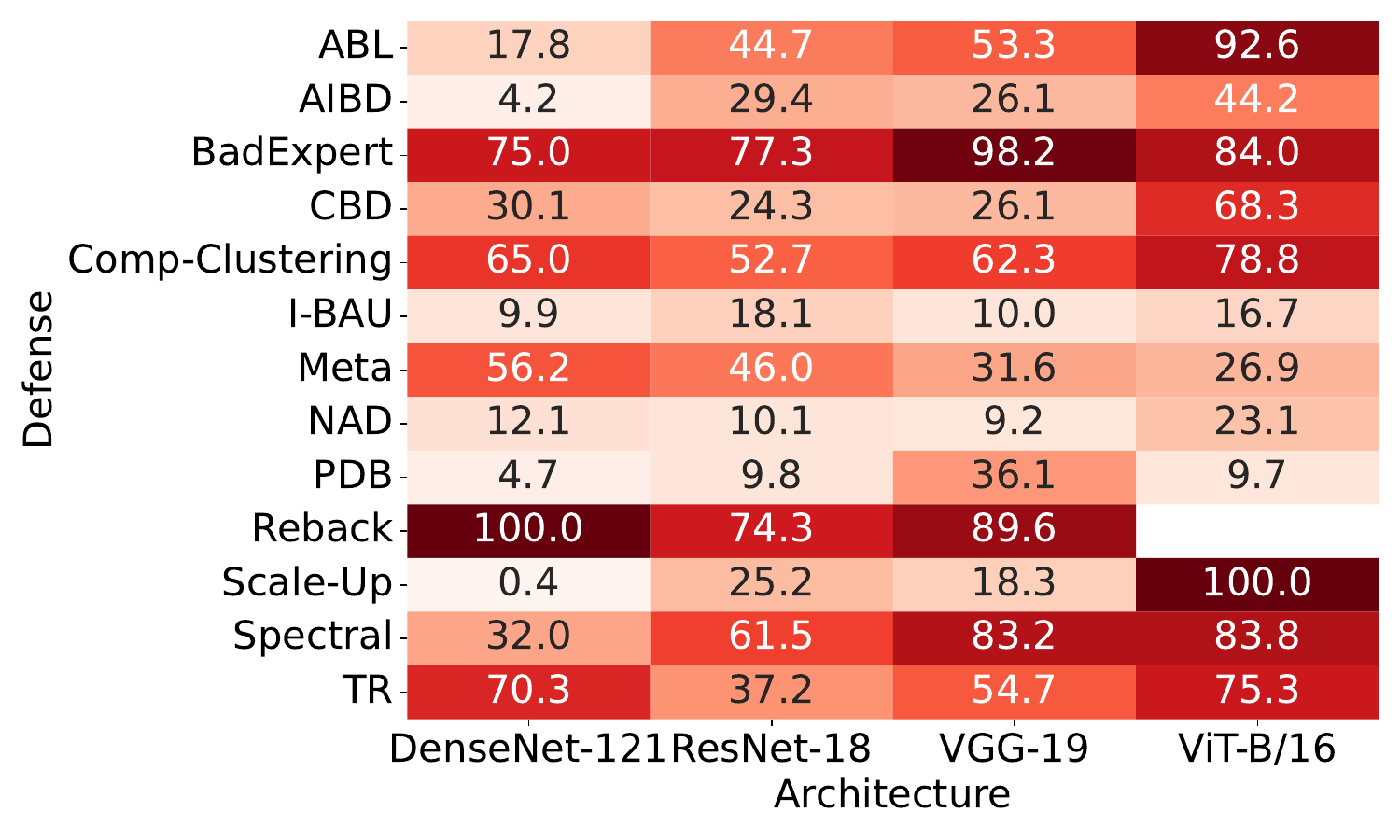}
        \caption{ASR (\%)}
    \end{subfigure}
    \hfill
    \begin{subfigure}[b]{0.8\linewidth}
        \centering
        \includegraphics[width=\linewidth, trim=10pt 13pt 10pt 10pt, clip]{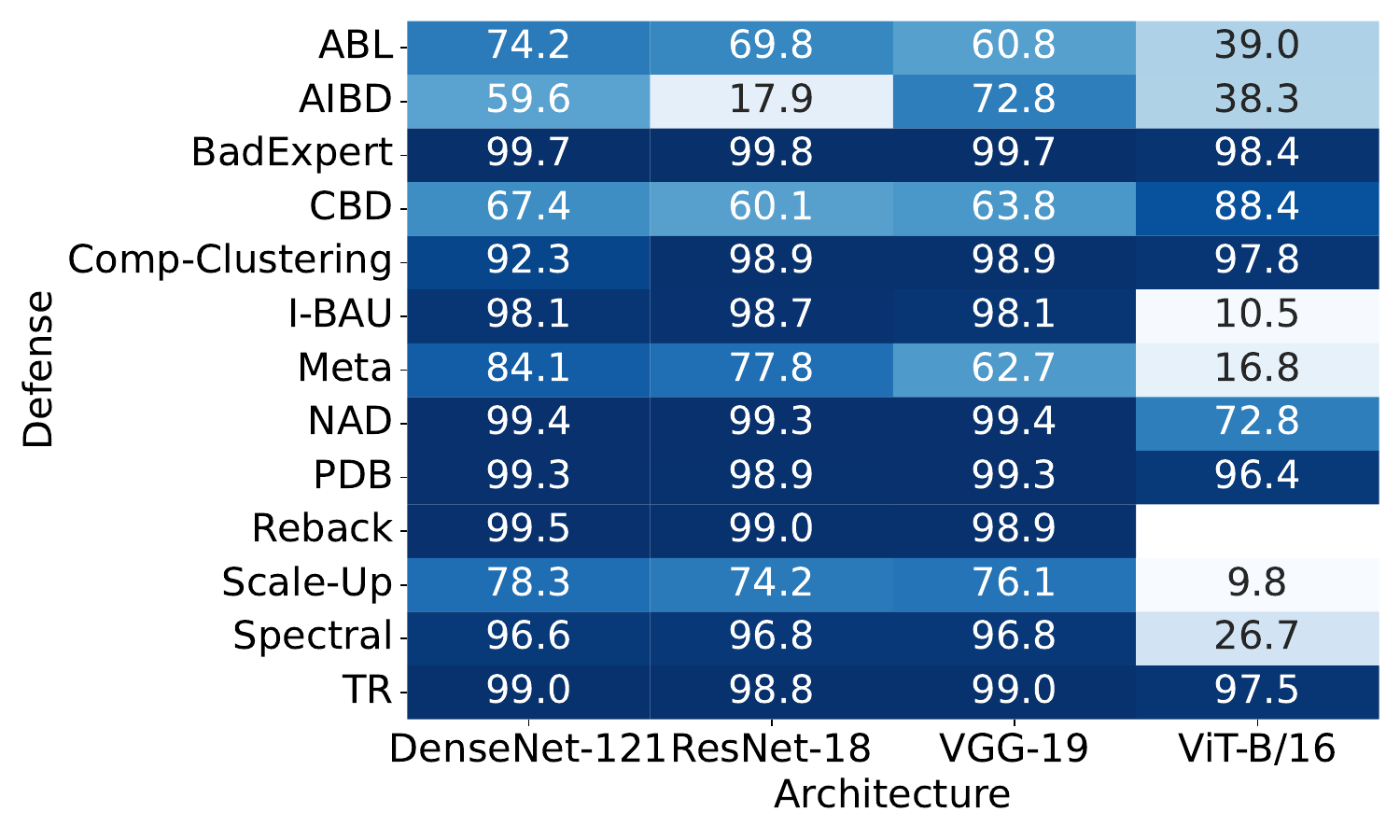}
        \caption{CA (\%)}
    \end{subfigure}
    \caption{Average CA and ASR for MNIST for different models.}
    \label{fig:MNISTAvgArchitecture}
\end{figure}

\begin{figure}[tbp]
    \centering
    \begin{subfigure}[b]{0.8\linewidth}
        \centering
        \includegraphics[width=\linewidth, trim=10pt 13pt 10pt 10pt, clip]{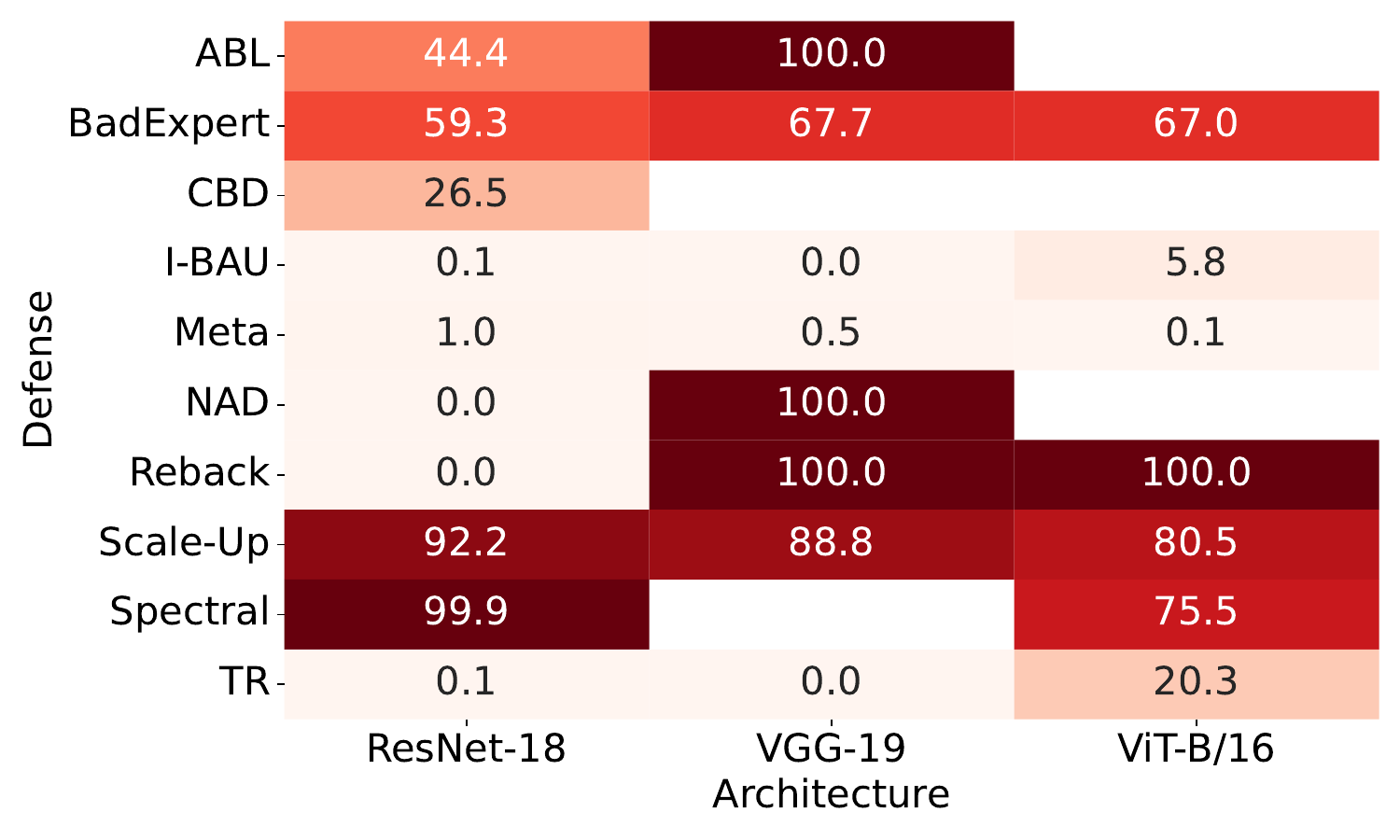}
        \caption{ASR (\%)}
    \end{subfigure}
    \hfill    
    \begin{subfigure}[b]{0.8\linewidth}
        \centering
        \includegraphics[width=\linewidth, trim=10pt 13pt 10pt 10pt, clip]{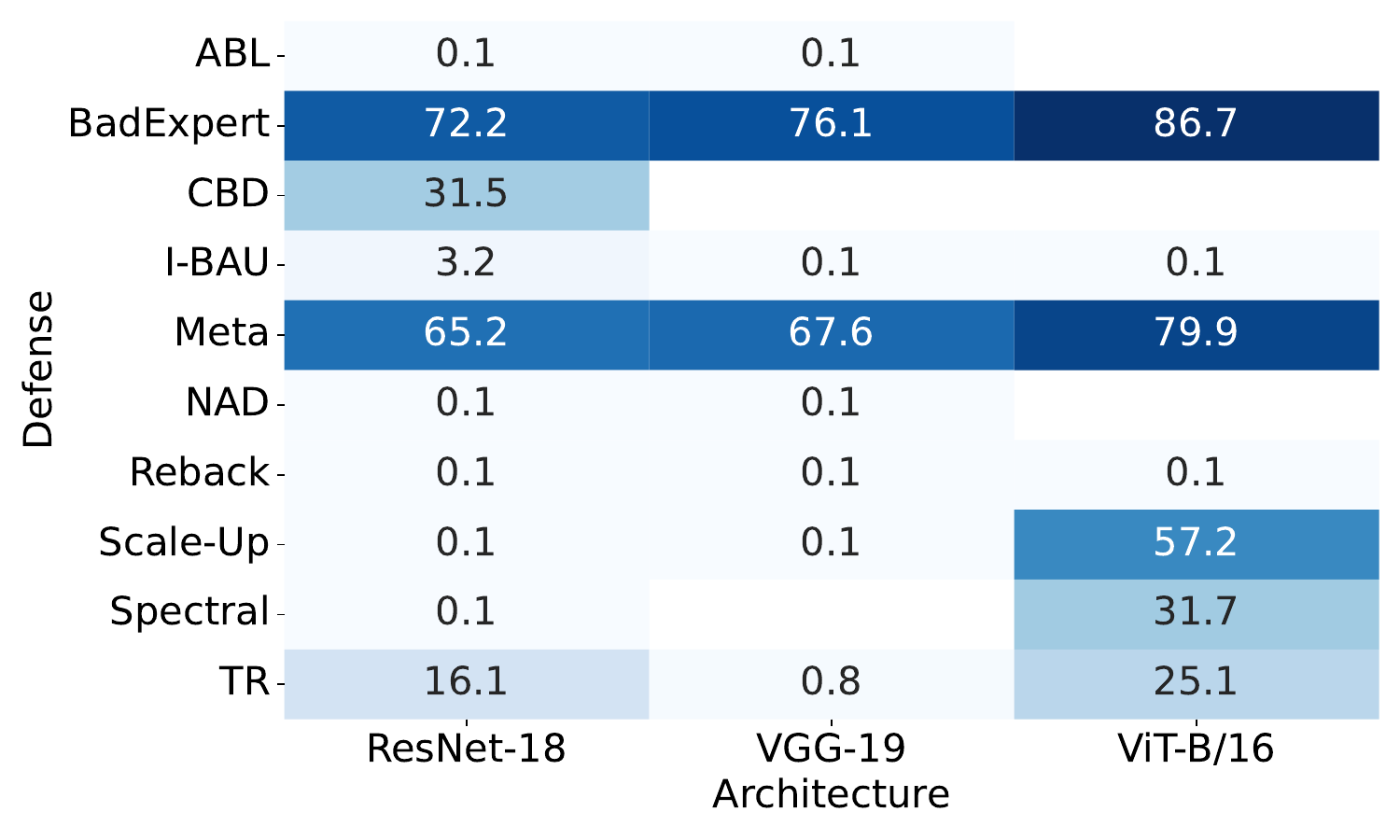}
        \caption{CA (\%)}
    \end{subfigure}
    \caption{Average CA and ASR for ImageNet-1K for different models.}
    \label{fig:ImagenetAvgArchitecture}
\end{figure}

\begin{figure}[tbp]
    \centering
    \begin{subfigure}[b]{0.8\linewidth}
        \centering
        \includegraphics[width=\linewidth, trim=10pt 13pt 10pt 10pt, clip]{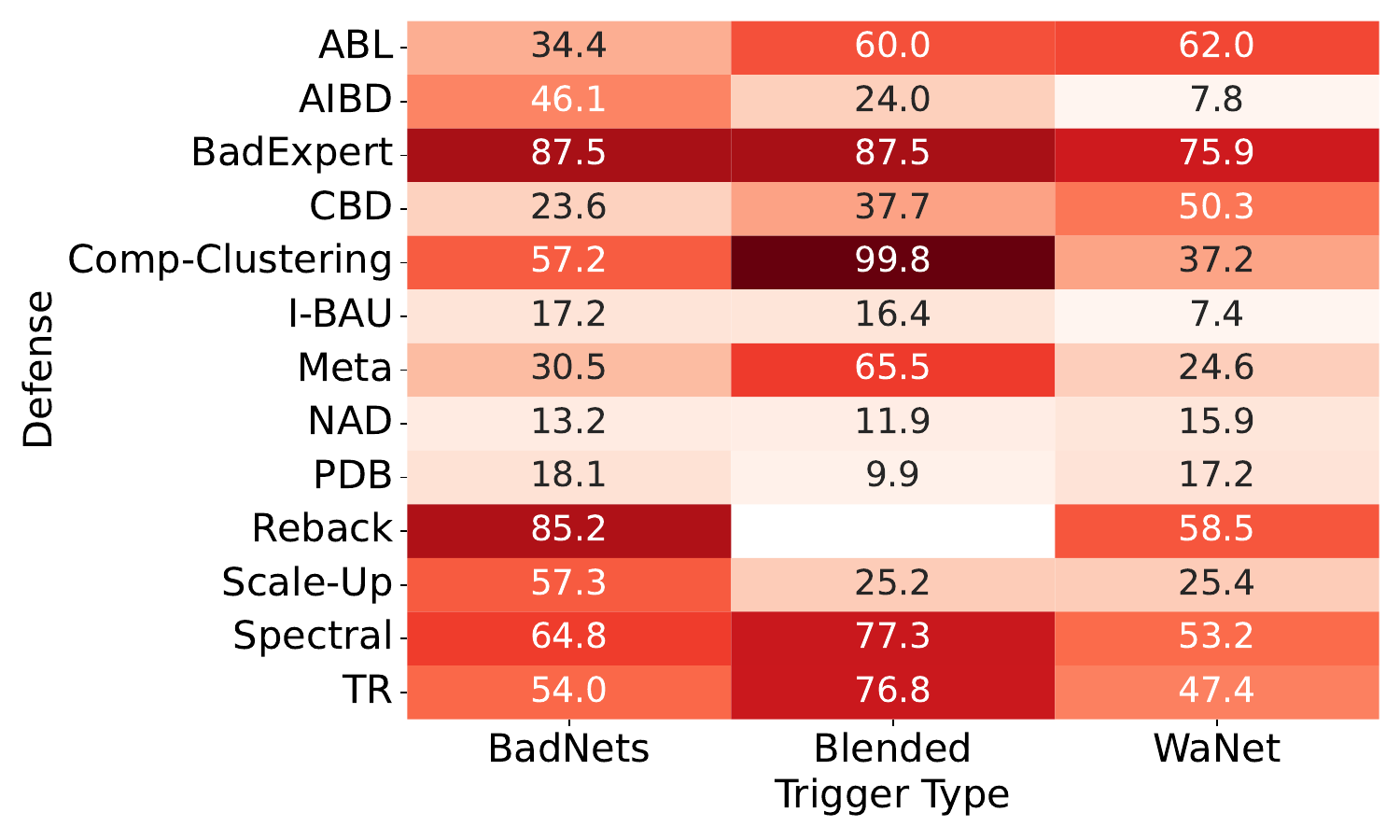}
        \caption{ASR (\%)}
    \end{subfigure}
    
    \begin{subfigure}[b]{0.8\linewidth}
        \centering
        \includegraphics[width=\linewidth, trim=10pt 13pt 10pt 10pt, clip]{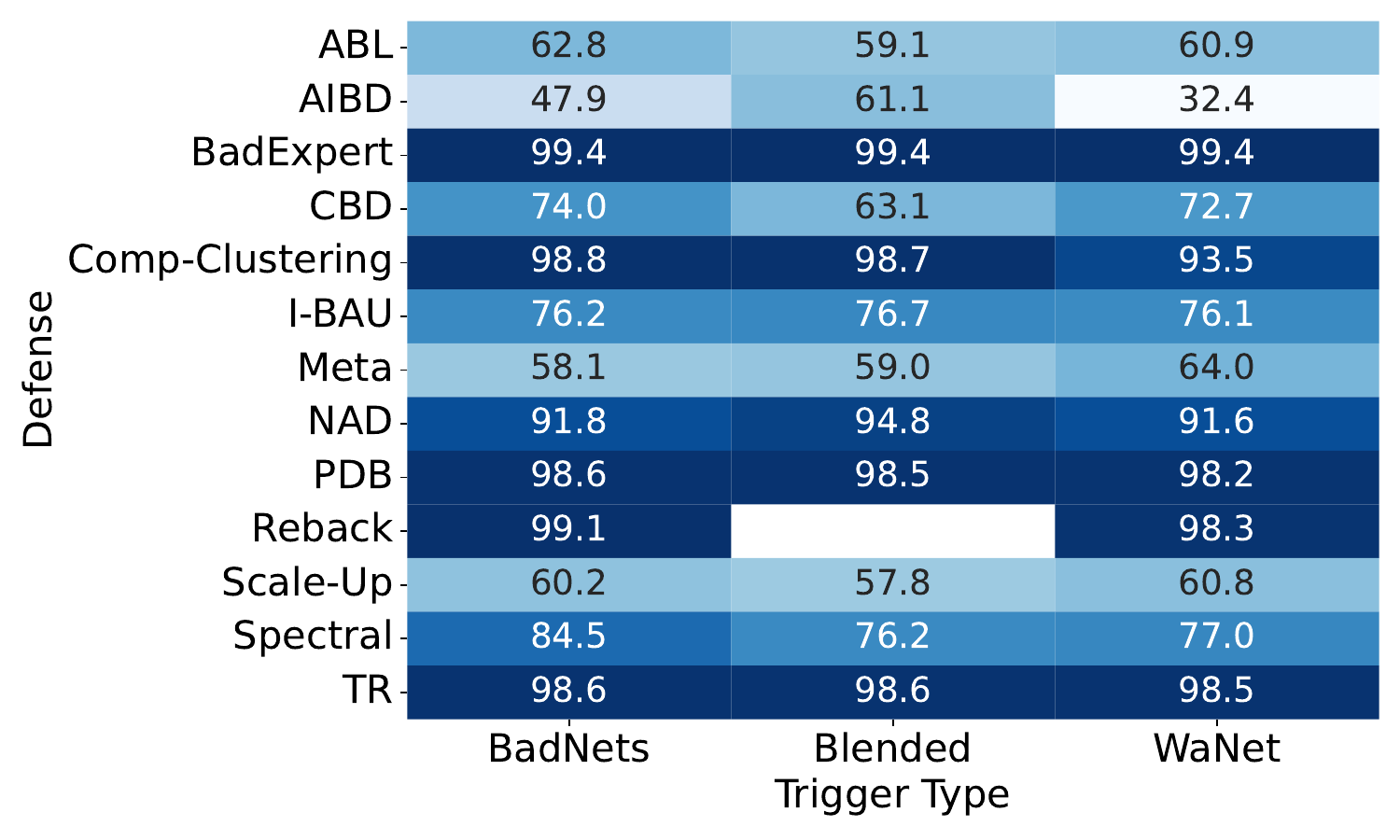}
        \caption{CA (\%)}
    \end{subfigure}
    \caption{Average CA and ASR for MNIST for different trigger types.}
    \label{fig:MNISTAvgTriggerType}
\end{figure}

\begin{figure}[tbp]
    \centering
    \begin{subfigure}[b]{0.8\linewidth}
        \centering
        \includegraphics[width=\linewidth, trim=10pt 13pt 10pt 10pt, clip]{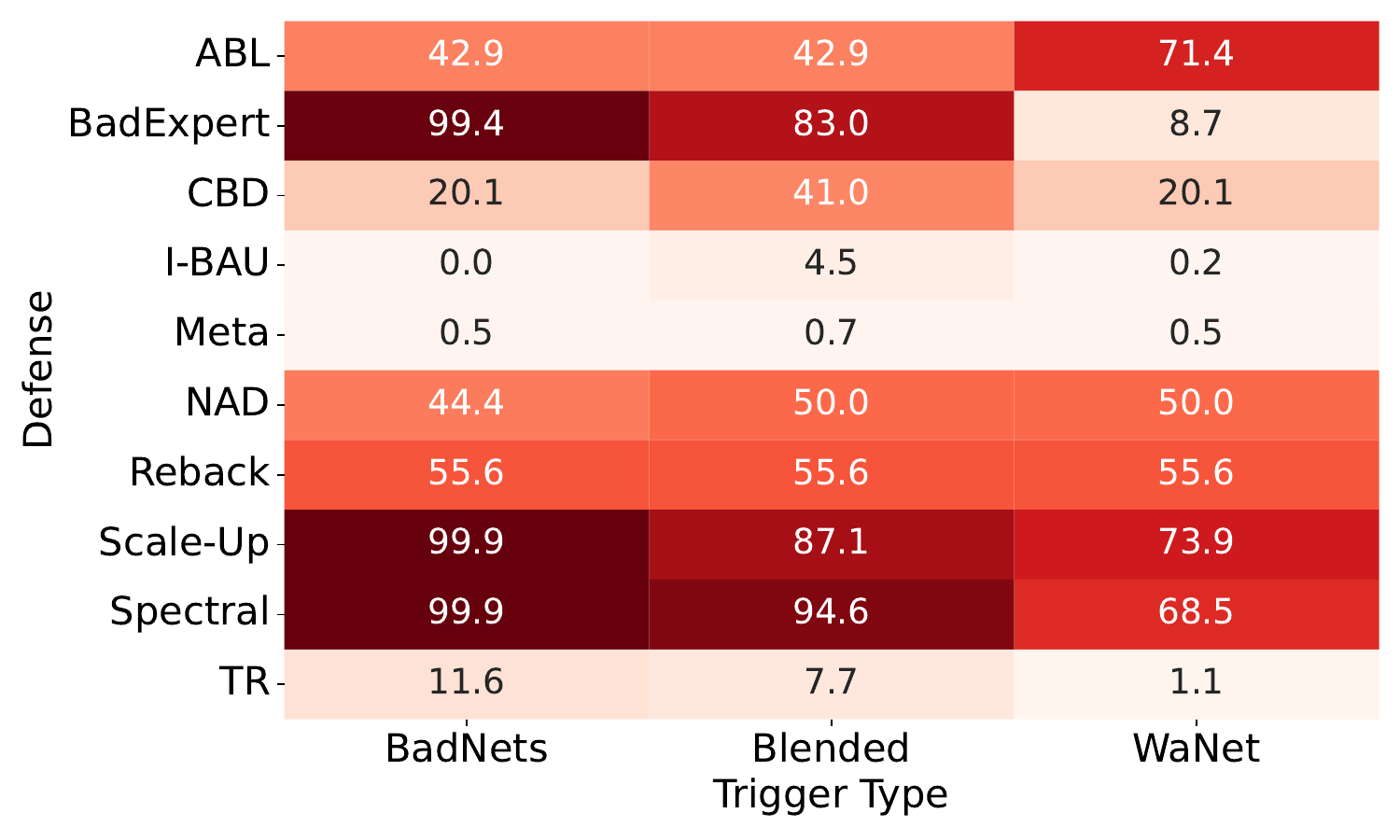}
        \caption{ASR (\%)}
    \end{subfigure}
    
    \begin{subfigure}[b]{0.8\linewidth}
        \centering
        \includegraphics[width=\linewidth, trim=10pt 13pt 10pt 10pt, clip]{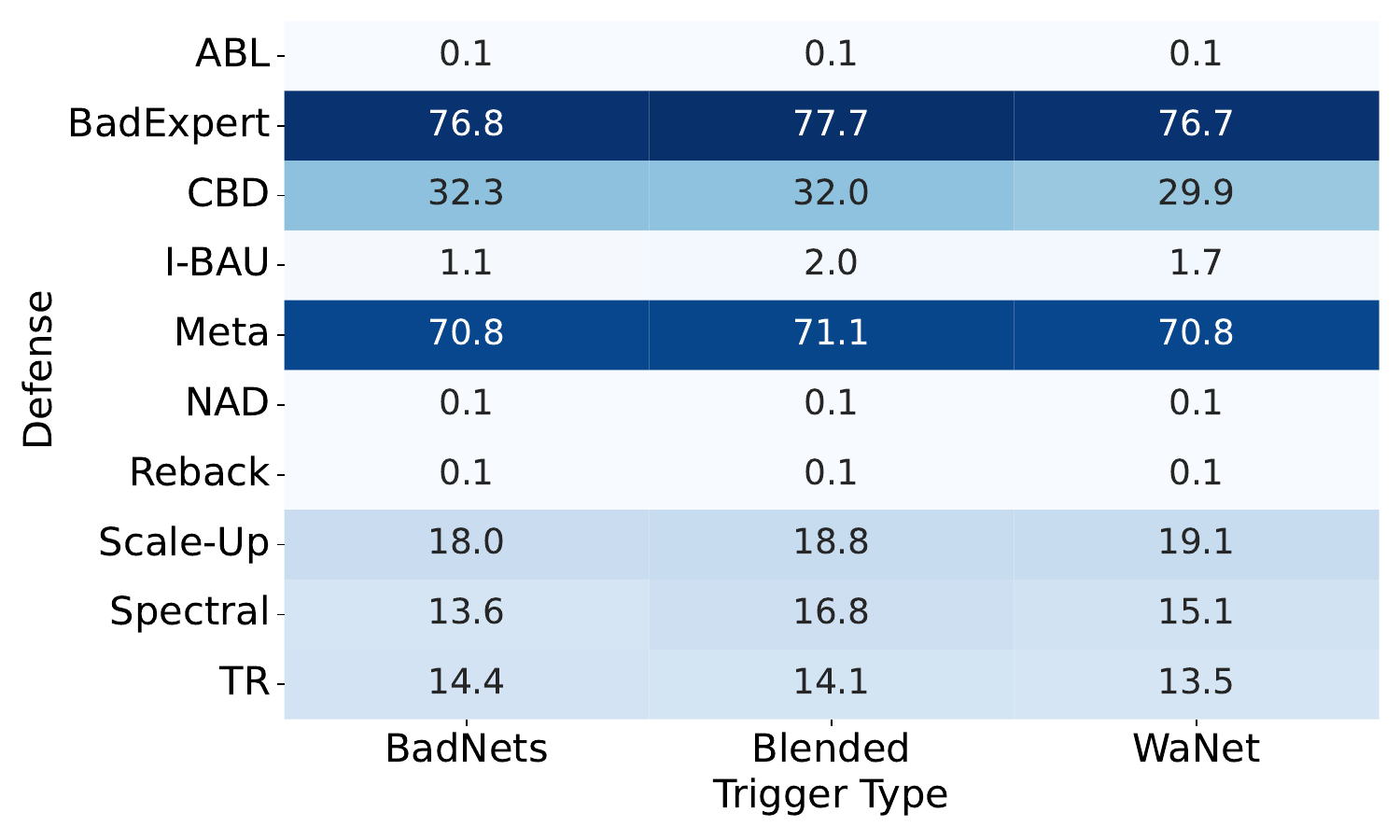}
        \caption{CA (\%)}
    \end{subfigure}
    \caption{Average CA and ASR for ImageNet-1K for different trigger types.}
    \label{fig:ImagenetTriggerType}
\end{figure}

\begin{figure}[tbp]
    \begin{subfigure}[b]{0.8\linewidth}
        \centering
        \includegraphics[width=\linewidth, trim=10pt 13pt 10pt 10pt, clip]{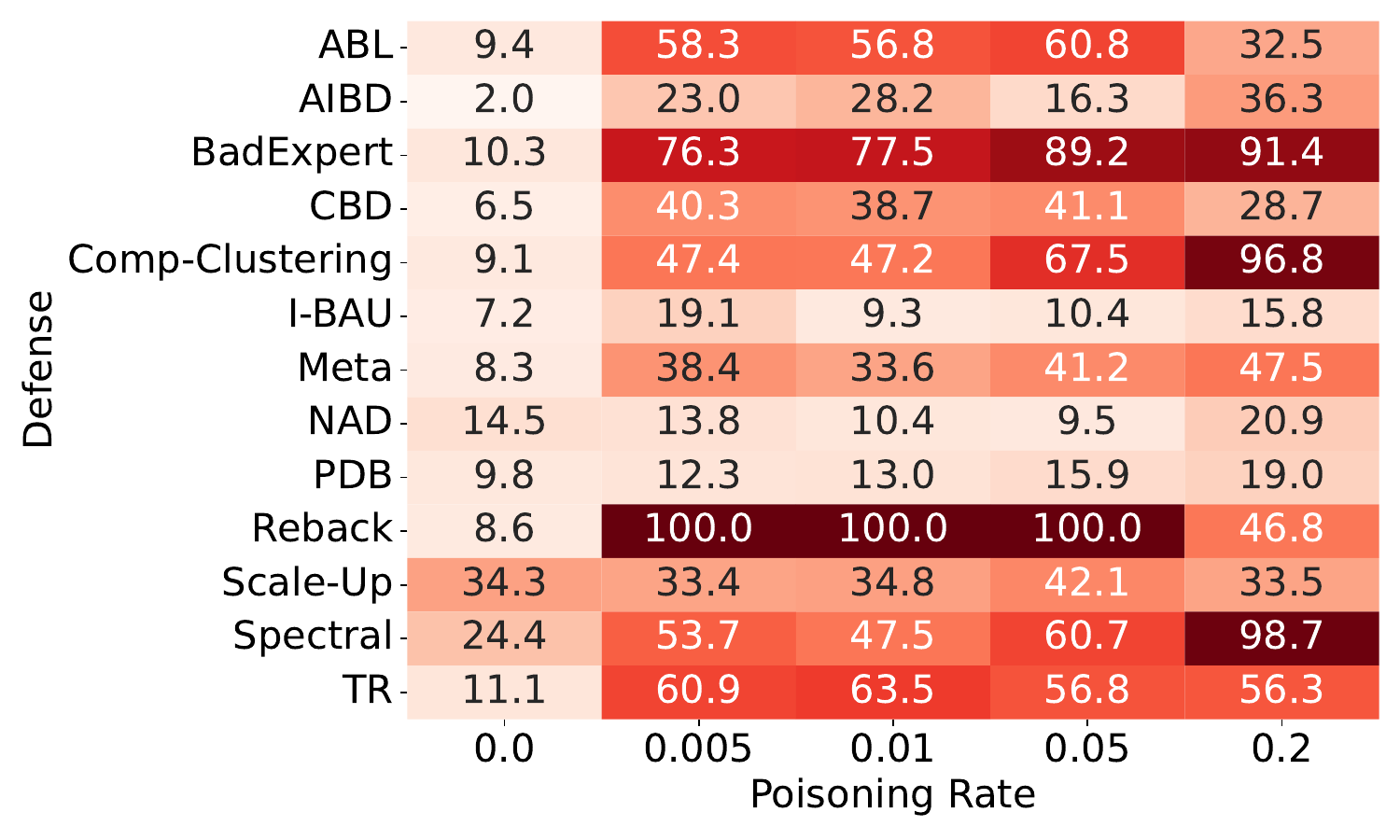}
        \caption{ASR (\%)}
    \end{subfigure}
    
    \begin{subfigure}[b]{0.8\linewidth}
        \centering
        \includegraphics[width=\linewidth, trim=10pt 13pt 10pt 10pt, clip]{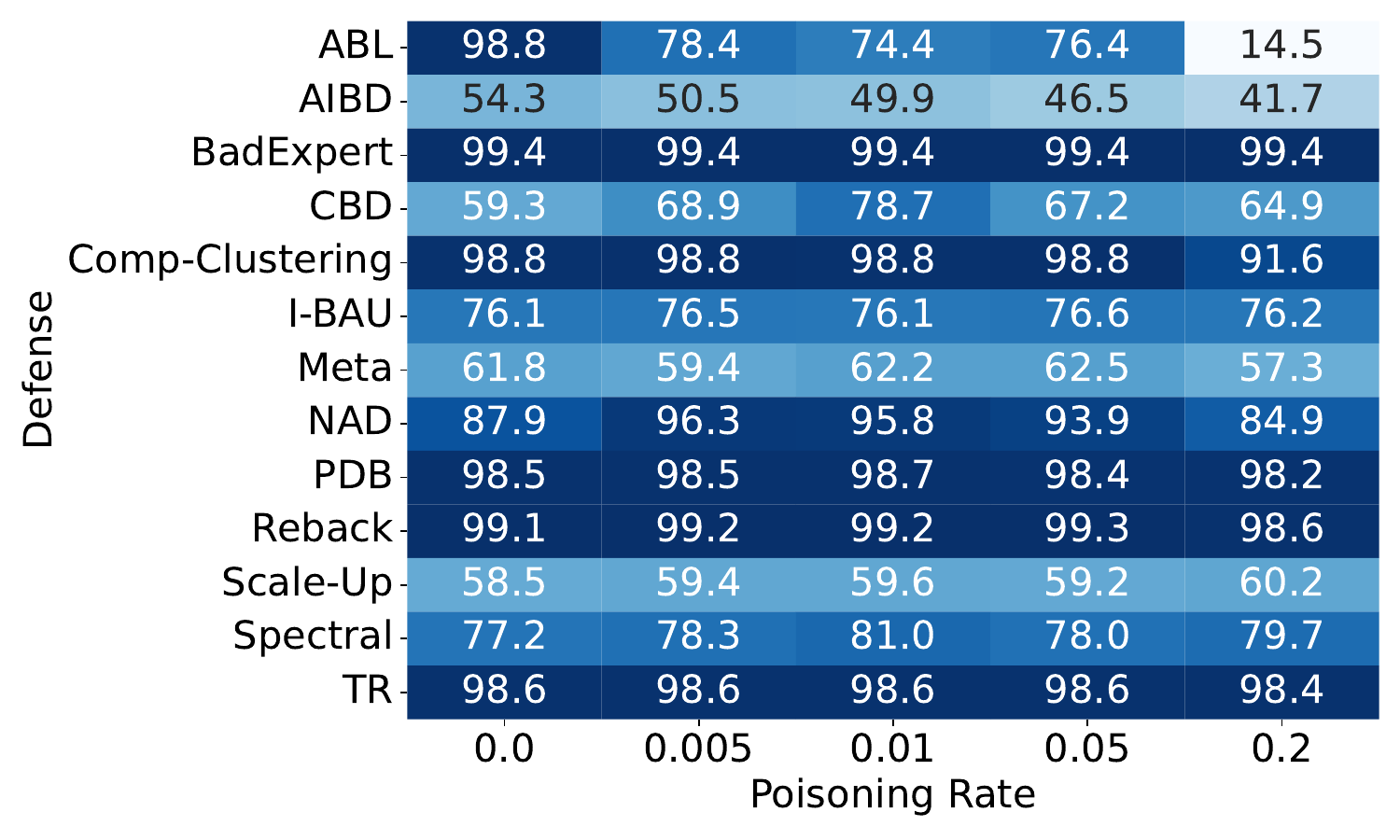}
        \caption{CA (\%)}
    \end{subfigure}
    \caption{Average CA and ASR for MNIST for different poisoning rates.}
    \label{fig:MNISTAvgPoisningRate}
\end{figure}

\begin{figure}[tbp]
    \begin{subfigure}[b]{0.8\linewidth}
        \centering
        \includegraphics[width=\linewidth, trim=10pt 13pt 10pt 10pt, clip]{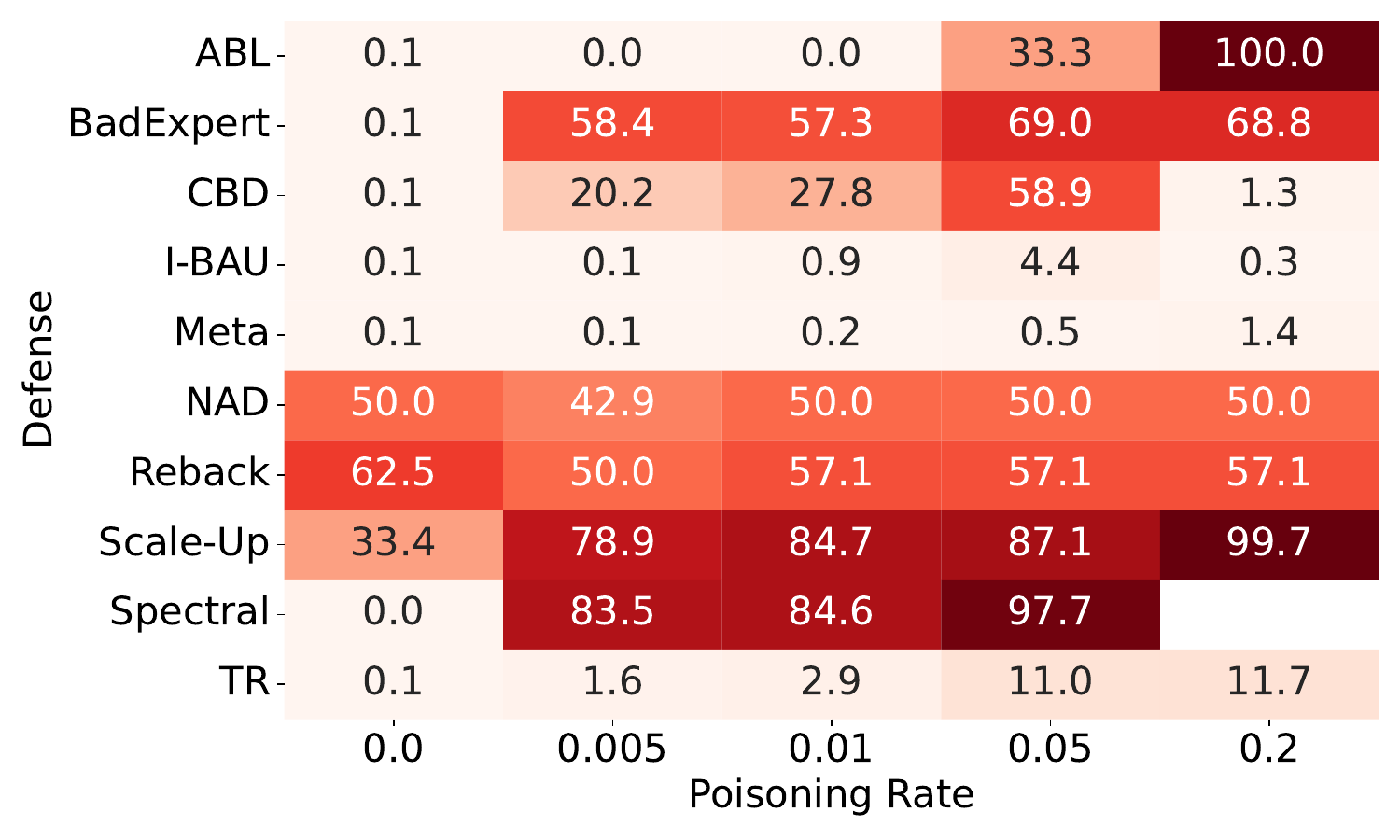}
        \caption{ASR (\%)}
    \end{subfigure}
    
    \begin{subfigure}[b]{0.8\linewidth}
        \centering
        \includegraphics[width=\linewidth, trim=10pt 13pt 10pt 10pt, clip]{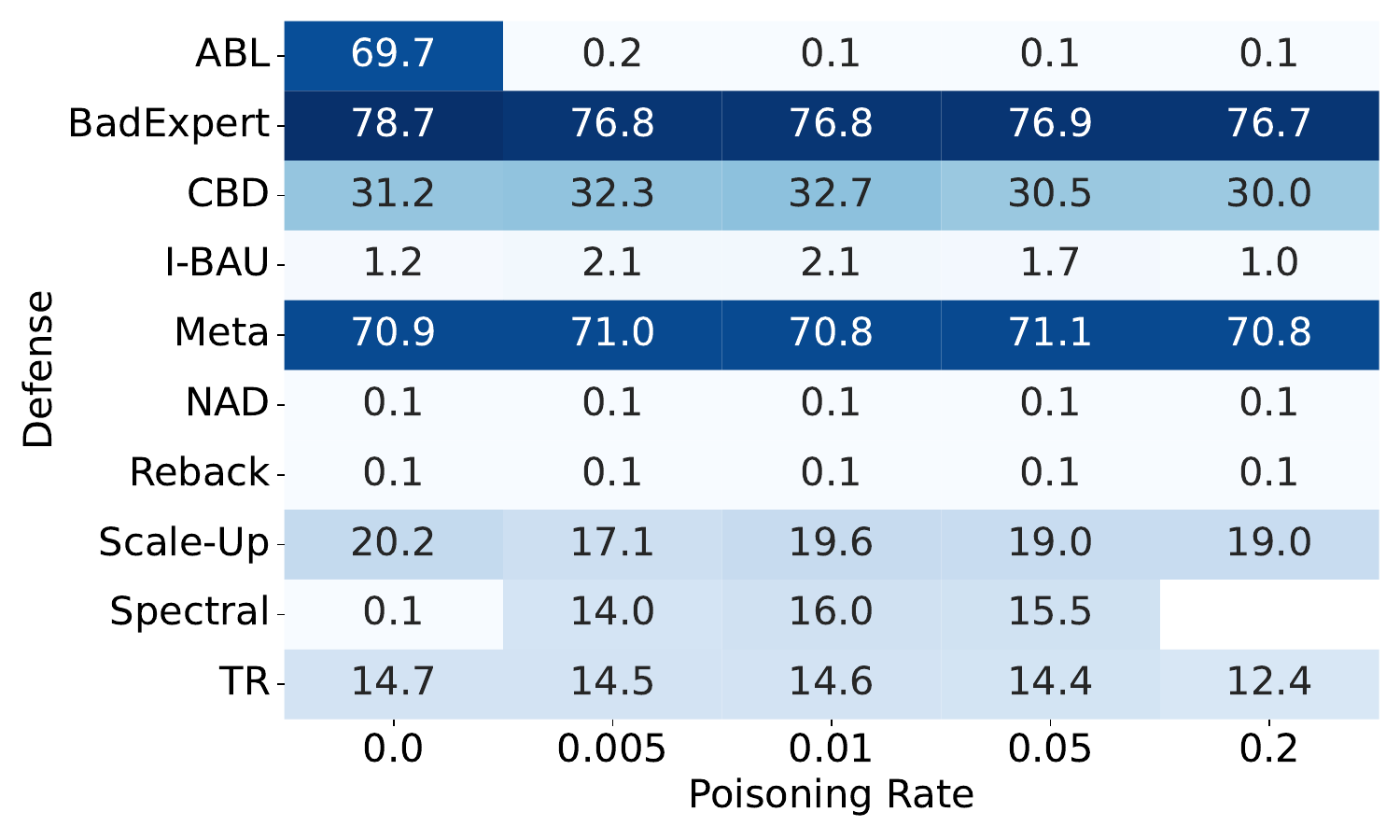}
        \caption{CA (\%)}
    \end{subfigure}
    \caption{Average CA and ASR for ImageNet-1K for different poisoning rates.}
    \label{fig:ImagenetAvgPoisningRate}
\end{figure}

Figures~\ref{fig:MNISTAvgArchitecture}, \ref{fig:MNISTAvgTriggerType}, \ref{fig:MNISTAvgPoisningRate}, and Figures~\ref{fig:ImagenetAvgArchitecture}, \ref{fig:ImagenetTriggerType}, \ref{fig:ImagenetAvgPoisningRate} show the performance of defenses on MNIST and ImageNet-1K datasets, respectively. Note that some results (like the defense performance with the Narcissus or AdvDoor attacks) are omitted due to the incompatibility or ineffective attack in the given scenarios.

\newpage
\bibliographystyle{ieeetr}
\bibliography{references}

\end{document}